\documentclass{sig-alternate-10pt}
\usepackage{verbatim,cite,url,listings,subfigure,times,setspace,psfrag,color,hyperref}
\usepackage{algorithm,algorithmic,multirow}

\usepackage[T1]{fontenc}
\usepackage{textcomp}
\usepackage[top=0.875in,bottom=0.875in,left=0.755in,right=0.755in,letterpaper]{geometry}

\begin{document}

\title{TouchIn: Sightless Two-factor Authentication on Multi-touch Mobile Devices}

\numberofauthors{4}
\author{
\alignauthor
Jingchao Sun \\
       \affaddr{Arizona State University}\\
       \email{jcsun@asu.edu}
\alignauthor
Rui Zhang \\
       \affaddr{University of Hawaii}\\
       \email{ruizhang@hawaii.edu}
\and
\alignauthor
Jinxue Zhang\\
       \affaddr{Arizona State University}\\
       \email{jxzhang@asu.edu}
\alignauthor Yanchao Zhang\\
       \affaddr{Arizona State University}\\
       \email{yczhang@asu.edu}
}

\maketitle
\begin{abstract}
Mobile authentication is indispensable for preventing unauthorized access to multi-touch mobile devices. Existing mobile authentication techniques are often cumbersome to use and also vulnerable to shoulder-surfing and smudge attacks. This paper focuses on designing, implementing, and evaluating TouchIn, a two-factor authentication system on multi-touch mobile devices. TouchIn works by letting a user draw on the touchscreen with one or multiple fingers to unlock his mobile device, and the user is authenticated based on the geometric properties of his drawn curves as well as his behavioral and physiological characteristics. TouchIn allows the user to draw on arbitrary regions on the touchscreen without looking at it. This nice sightless feature makes TouchIn very easy to use and also robust to shoulder-surfing and smudge attacks. Comprehensive experiments on Android devices confirm the high security and usability of TouchIn.
\end{abstract}

\section{Introduction}\label{sec:Intro}
Recent years have seen the explosive growth of multi-touch mobile devices such as iPads and Nexus~7 tablets. In computing, multi-touch refers to the ability of sensing the input from two or more points of contact with a touchscreen at one time. The great demand for multi-touch technology is largely driven by the skyrocketing quest for multi-touch smartphones, tablets, and other consumer electronic devices.

Mobile (user) authentication---letting multi-touch mobile devices ascertain whom they are interacting with---is necessary for preventing unauthorized access to mobile devices which store increasingly more private information. Existing mobile authentication techniques can be broadly classified into three paradigms: \emph{something you know} such as alphanumeric/graphical passwords, \emph{something you have} such as a hardware token \cite{BojinMob11,VuDis12,NichoMob06,RothIr10}, and \emph{someone you are} such as biological and behavioral characteristics \cite{KurkoExp10,DerawFin11,ShiImp10,MantyIde05,GafurSpo07,MaiorKey11,LucaTou12,SandnUse12,LiUno13}. Multi-factor mobile authentication refers to the reliance on more than one authentication paradigm.

The something-you-know paradigm has so far been the most widely used. For example, a simple password on iOS devices is a 4-digit number, and the Android Pattern Lock technique uses a graphical password as a pattern on a 9-point grid. To input a correct password, a user has to look at the screen and touch specific points on a virtual keyboard or grid. Such password entry has four essential drawbacks. First, it is cumbersome and a common frustration so that many (if not most) people just avoid password protection \cite{AzenkPas12,VuDis12}. Second, it is vulnerable to shoulder-surfing attacks because the password can be potentially observed by malicious bystanders in crowded public places \cite{SaeBio12,ShahzSec13}. Third, it is susceptible to smudge attacks \cite{AvivSmu10,ShahzSec13}, as the user repeatedly touches the same points on the screen before his password changes. Last, it is inaccessible to people with visual impairment \cite{AzenkPas12}, corresponding to 285 million people worldwide \cite{WhoBlind} and 21.5 million US adults aged 18+ \cite{USBlind}.

Although free of the above drawbacks, the other paradigms also have limitations. In particular, the something-you-have paradigm often requires specifically built auxiliary hardware such as hardware tokens \cite{BojinMob11,VuDis12} not immediately available on the market. In addition, belonging to the someone-you-are paradigm, biometric authentication techniques relying on face/iris/voice/fingerprint recognition are vulnerable to well-known spoofing mechanisms \cite{VuDis12}. For example, the fingerprint authentication feature on the latest iPhone 5S has been quickly broken \cite{Hackfingerprint,Howto}.

This paper presents the design, implementation, and evaluation of \textbf{TouchIn}, a two-factor authentication system for multi-touch mobile devices as a novel solution to the aforementioned issues. Our major contributions are twofold.

First, we design and implement TouchIn as a novel combination of the something-you-know and someone-you-are paradigms. TouchIn comprises two phases. In the \emph{enrollment} phase, a device owner uses one or multiple fingers to draw arbitrary geometric curves of his own choice (called a \emph{curve password}) on his multi-touch screen. An \emph{authentication template} is then created based on features extracted from his input, including x-coordinate, y-coordinate, direction, curvature, x-velocity, y-velocity, x-acceleration, y-acceleration, finger pressure, and hand geometry. The first four features relate to the something-you-know paradigm and can together accurately define the geometric characteristics of the drawn curve. They are incorporated based on the observation that self-defined curves are easier for the user himself to remember and reproduce but difficult for attackers to guess. In contrast, associated with the something-you-are paradigm, the remaining six features correspond to the user's behaviorial characteristics and physiological characteristics. They are nearly impossible for attackers to infer and forge even if they may manage to know the correct curve password. We formulate the weighted combination of these features as an optimization problem and propose a solution based on machine learning. In the subsequent \emph{authentication phase}, anyone attempting to unlock the mobile device needs to draw on the multi-touch screen, from which a \emph{candidate template} is extracted. If the candidate and authentication templates match, the user is allowed in.

Second, we implement TouchIn on Google Nexus 7 tablets and evaluate its security and usability through comprehensive experiments under various adversary models. Specifically, our security and usability studies involve 30 volunteers with their eyes closed throughout the experiments to emulate a rather constrained environment where TouchIn is used by people with visual impairments. We show that TouchIn has very low false positives and negatives, denying unauthorized users for 97.7\% (97.8\%) of the time and admitting authorized users for 97.5\% (99.3\%) of the time for a password involving a single curve (multiple curves). Finally, our usability survey shows that most volunteers find TouchIn easy to use and curve passwords easy to memorize.

TouchIn does not suffer from the aforementioned drawbacks observed with existing mobile authentication techniques. First, it is a \emph{sightless} solution in that the user can draw his password curve without seeing the screen. There are two major implications from this feature: (1) TouchIn is both accessible and user-friendly for sighted people and also people with visual impairment; (2) the user can perform device unlocking under some cover (say a briefcase or jacket) to completely thwart shoulder-surfing attacks \cite{SaeBio12,ShahzSec13}. Second, the user can draw anywhere on the touchscreen under an arbitrary orientation for each unlocking attempt. Finger smudges can thus be more randomly distributed over a larger portion of the screen instead of at some fixed points, making smudge attacks \cite{AvivSmu10,ShahzSec13} much less a threat. Third, TouchIn does not require any auxiliary hardware and is applicable to off-the-shelf multi-touch mobile devices. Finally, TouchIn is highly secure due to its reliance on two authentication paradigms.

The rest of this paper is organized as follows. Section~\ref{sec:Related} surveys the related work. Section~\ref{sec:Basics} introduces the multi-touch basics. Section~\ref{sec:Adversary} presents the adversary model. Section~\ref{Sec:DataAnalysis} analyzes some collected data to motivate our design. Section~\ref{sec:DesignOverview} overviews the TouchIn design. Section~\ref{Sec:Enrollment} details the enrollment phase of TouchIn. Section~\ref{Sec:Verification} illustrates the verification phase of TouchIn. Section~\ref{sec:Experiment} evaluates the TouchIn performance through comprehensive experiments. Section~\ref{sec:Conclusion} concludes this paper.
\vspace{.1in}

\section{Related Work}\label{sec:Related}
Due to space limitations, we only brief the prior work most germane to our TouchIn system.

In addition to alphanumeric and graphic passwords, gesture passwords are associated with the something-you-know paradigm. Bailador \emph{et al.} \cite{BailaAna07} proposed to authenticate a mobile user by letting him make a handwritten signature in the air while holding a mobile phone, and the in-air signature is captured through the 3D accelerometer sensor pervasive on smartphones and tablets nowadays for comparison with one previously stored on the phone. A similar idea is presented in \cite{LiuUse09,LiuUwa09} to use the accelerometer sensor to capture user-created gestures. Besides being socially awkward, these techniques can only be used as weak authentication techniques and are
vulnerable to the attackers seeing the users perform their in-air signatures or gestures, as mentioned in \cite{BailaAna07,LiuUse09,LiuUwa09}. In \cite{SaeBio12}, a comprehensive set of five-finger gestures
are defined for multi-touch device authentication. The shape of each user performing a given gesture is used and evaluated as his device password. The security of this technique and the use of personalized gestures instead of
predefined ones are not fully investigated. In addition, performing five-finger gestures on mobile devices with a small display may not be user-friendly. Moreover, Kinwrite \cite{TianKin13} is a handwriting-based gesture authentication system, but it requires an external device like Kinect to detect 3D handwriting motions. Most recently, GEAT \cite{ShahzSec13} authenticates a mobile user based on how he inputs the specific gestures and does not fulfill the sightless requirement. Finally, PassChords \cite{AzenkPas12} is the only prior work dedicated to the sightless requirement to the best of our knowledge. It requires a user to tap several times on the multi-touch screen with one or more fingers, and the sets of fingers in all the taps together compose a password for comparison with one stored on the mobile device. Although promising results have been shown from usability studies, the authentication failure rate is 16.3\% \cite{AzenkPas12}, and our experiments reveal that PassChords is very vulnerable to shoulder-surfing attacks.

Online handwritten signature verification on mobile devices such as \cite{MendaAna10,MendaSec11} relates to the something-you-know paradigm as well. In such a system, a user uses a stylus to write a handwritten signature on the touchscreen, which is then captured and compared with a template on the device.
A handwritten signature can be regarded as a special single-curve password in TouchIn, but the majority of publicly reported systems along this line assume a stylus used for better handwritten signature recognition. In contrast, TouchIn supports both single-curve and multi-curve authentication, relies only on fingers for better usability, and incorporates a user's behavioral and physiological characteristics as well.

The something-you-have authentication paradigm often requires special auxiliary devices not immediately available on the market. Knuepfel made Signet Rings that use conductive material to create several electrical pathways from a carrier's fingers
to the capacitive touchscreen, and the pathways are arranged in a distinct pattern as a password \cite{Signet11}. A similar technique with
more details is introduced in \cite{VuDis12}. In addition, the Magkey and Mickey miniature devices presented in \cite{BojinMob11} rely on secret-embedded magnet fields and acoustic signals detected by the smartphone's
compass sensor and microphone, respectively. Furthermore, a wearable token is used in \cite{NichoMob06} to keep attesting the user's presence to his mobile device by sending authentication messages via a short-range
wireless link (e.g., Bluetooth). Finally, the IR Ring in \cite{RothIr10} authenticates a user's touches to a multi-touch display by transmitting a cryptographic signal through the infrared channel. In contrast, TouchIn is for off-the-shelf multi-touch mobile devices and requires no special auxiliary devices.

The someone-you-are authentication paradigm usually depends on physiological or behavioral biometrics.
Physiological biometrics relate to a person's physical features such as fingerprints, iris patterns, retina patterns, facial features, palm prints, hand geometry. These features are difficult to be accurately identified
on mobile devices and also susceptible to well-known spoofing mechanisms \cite{FaundOn04,VuDis12}. In contrast, behavioral biometrics relate to a user's behavioral patterns such as location traces
\cite{ShiImp10}, gaits \cite{MantyIde05,GafurSpo07}, keystroke patterns \cite{MaiorKey11}, and touch dynamics \cite{LucaTou12,LiUno13}. These techniques are best suitable as secondary
authentication mechanisms supplementing the primary password-based authentication mechanism, as they may be vulnerable to attackers familiar with the
victim's behavioral patterns \cite{ShiImp10,LucaTou12}.
\vspace{-0.1in}

\section{Multi-touch Basics}\label{sec:Basics}
In this section, we outline the fundamentals of multi-touch screens to help understand TouchIn.

The history of multi-touch technology dated back to 1982, when the University of Toronto's Input Research Group developed the first human-input multi-touch system. A multi-touch screen is rapidly becoming commonplace on most recently shipped smartphones and tablets. Multi-touch can be implemented in different ways such as capacitive, resistive, optical, and wave technologies. We focus on the most popular capacitive multi-touch screens. Our security and usability studies involve popular Android devices. When a user draws with one or multiple fingers on the touchscreen, every finger will trigger a sequence of touch events which can be retrieved from Android OS. Every touch event can be characterized by a set of features, among which the following are relevant to TouchIn: \emph{finger ID} assigned to and uniquely identifying every finger during the finger motion, \emph{coordinate} of the touch point, \emph{timestamp} of the touch event, \emph{pressure} and \emph{size} applied to the touchscreen. For the Cartesian touch coordinates, the origin is at the top-left corner of the touchscreen, and the left-to-right and top-to-down directions define the \emph{x}-axis and \emph{y}-axis directions, respectively. Note that Android devices can support both landscape and portrait orientations, in which case the origin will vary. To avoid the confusion, the touchscreen is automatically locked to the portrait orientation during authentication. In addition, the pressure and size range from zero to one in Android OS.
\vspace{-0.1in}

\section{Adversary Model}\label{sec:Adversary}
As discussed, our TouchIn system relies on the device owner's self-defined geometric curves, which are referred to as the \emph{curve password} hereafter. Motivated by \cite{TianKin13}, we consider four adversary models from the weakest to strongest.

\begin{itemize}
  \item \textbf{Type-I}: The adversary knows neither the shapes of the curve password nor how the device owner draws the curves to unlock the mobile device.

  \item \textbf{Type-II}: The adversary can observe how the device owner draws the curves but not the curve shapes.

  \item \textbf{Type-III}: The adversary can observe how the device owner draws the curves and also the rough curve shapes. For example, the adversary may know that the curve password is an English letter but not the exact shape.

  \item \textbf{Type-IV}: The adversary knows exactly how the device owner draws the curves and also the curve shapes.
\end{itemize}

\section{Analysis of User-drawn Curves (or Curve Feature Selection)}\label{Sec:DataAnalysis}
To motivate the TouchIn design, here we analyze many collected user-drawn curves to show that there are extractable features fairly consistent for each individual user and also distinguishable among different users.

\subsection{Data Collection}\label{subsec:DataCollection}
We recruited 30 volunteers (six female and 24 male) over a two-week period, all of whom are aged 20 to 30 and pursuing their BE/MS/PhD degrees in Electrical Engineering or Computer Science.\footnote{The data collection process has gone through the IRB approval.} The experiments were done on two Google Nexus 7 tablets with 1.6 GHz NVIDIA Tegra 3 quad-core processor, 1G RAM, 16G internal storage, a 7-inch capacitive multi-touch screen, and Android 4.2 OS. Each volunteer was first briefed about 10 minutes for our research studies and was asked to close his/her eyes while drawing on the touchscreen to emulate the sightless requirement. In addition, each volunteer was asked to independently decide three different curves with each involving one finger (single-curve for short) and three others with each involving multiple fingers of the same hand (multi-curve for short). Finally, each volunteer was instructed to draw each of his/her own six curves 20 to 50 times, leading to 3600 single-curve samples and 3600 multi-curve samples.
\begin{figure}[!t]
\centering
    \subfigure[x-coordinate]{
        \label{xposition}
        \includegraphics[width=0.225\textwidth]{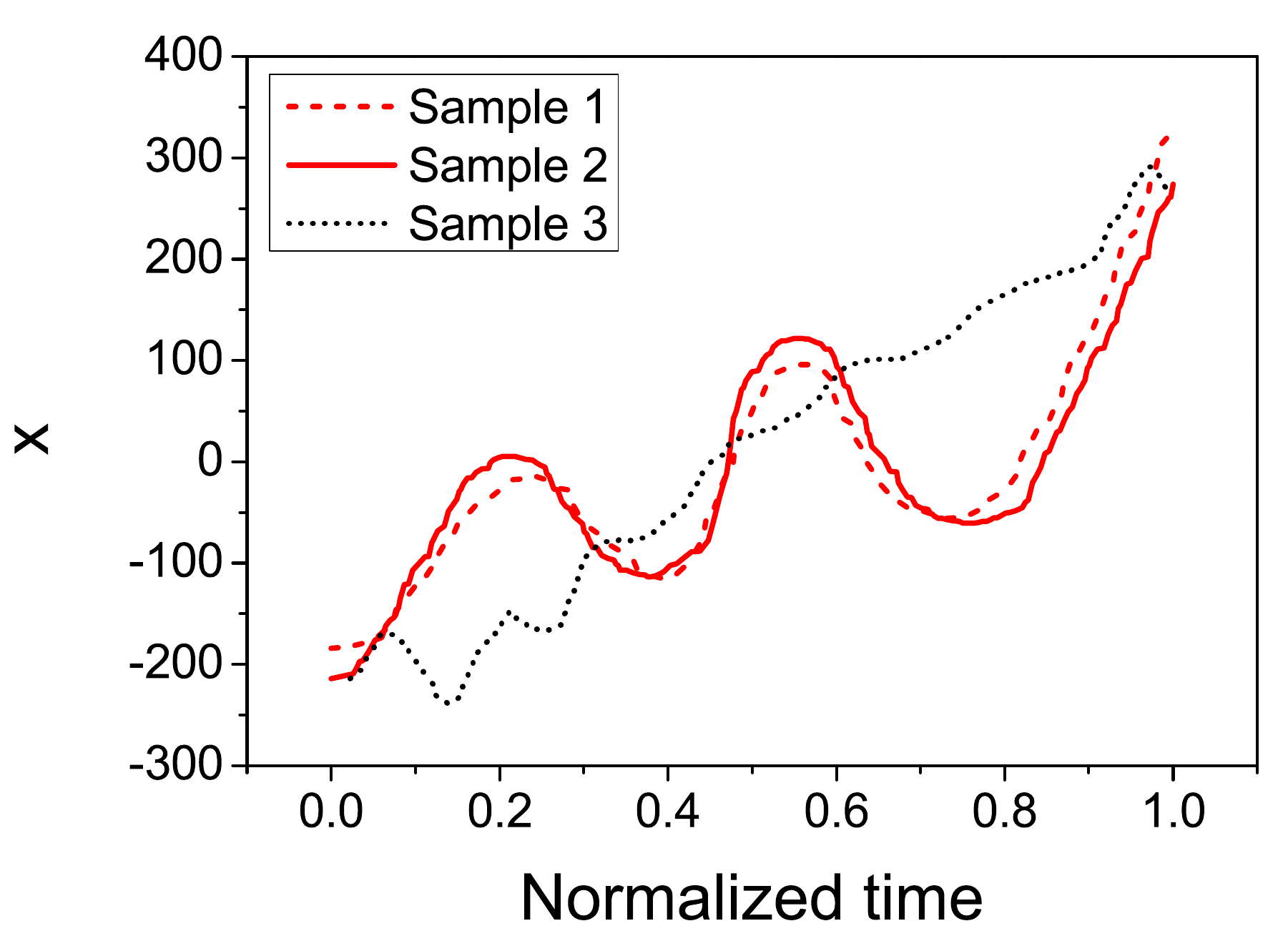}
    }
\hfill
    \subfigure[y-coordinate]{
        \label{yposition}
        \includegraphics[width=0.225\textwidth]{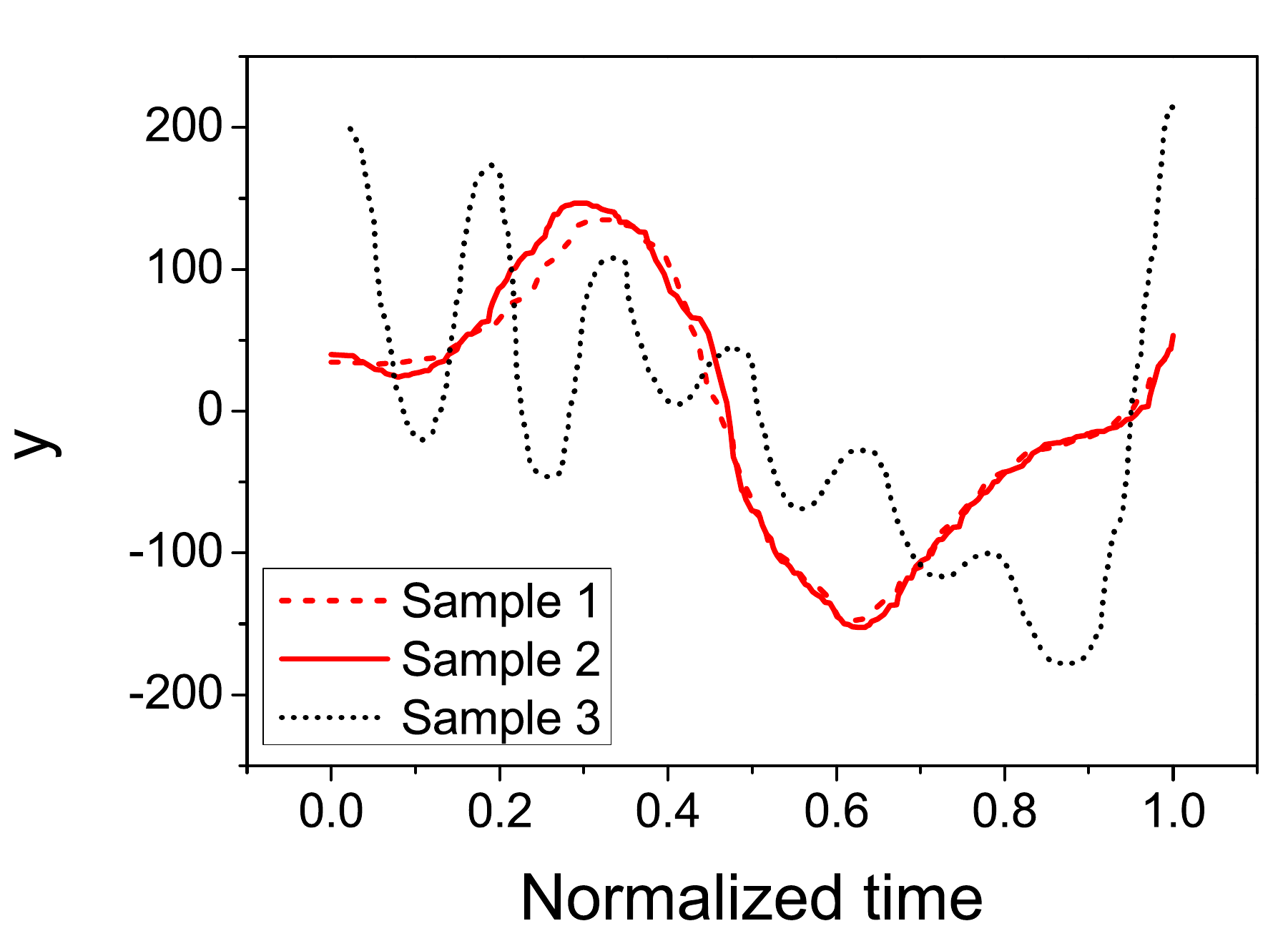}
    }
\caption{Coordinates of touch points.}\label{position}
\end{figure}

\begin{figure}[!t]
\centering
    \subfigure[Velocity in x direction]{
        \label{xvelocity}
        \includegraphics[width=0.225\textwidth]{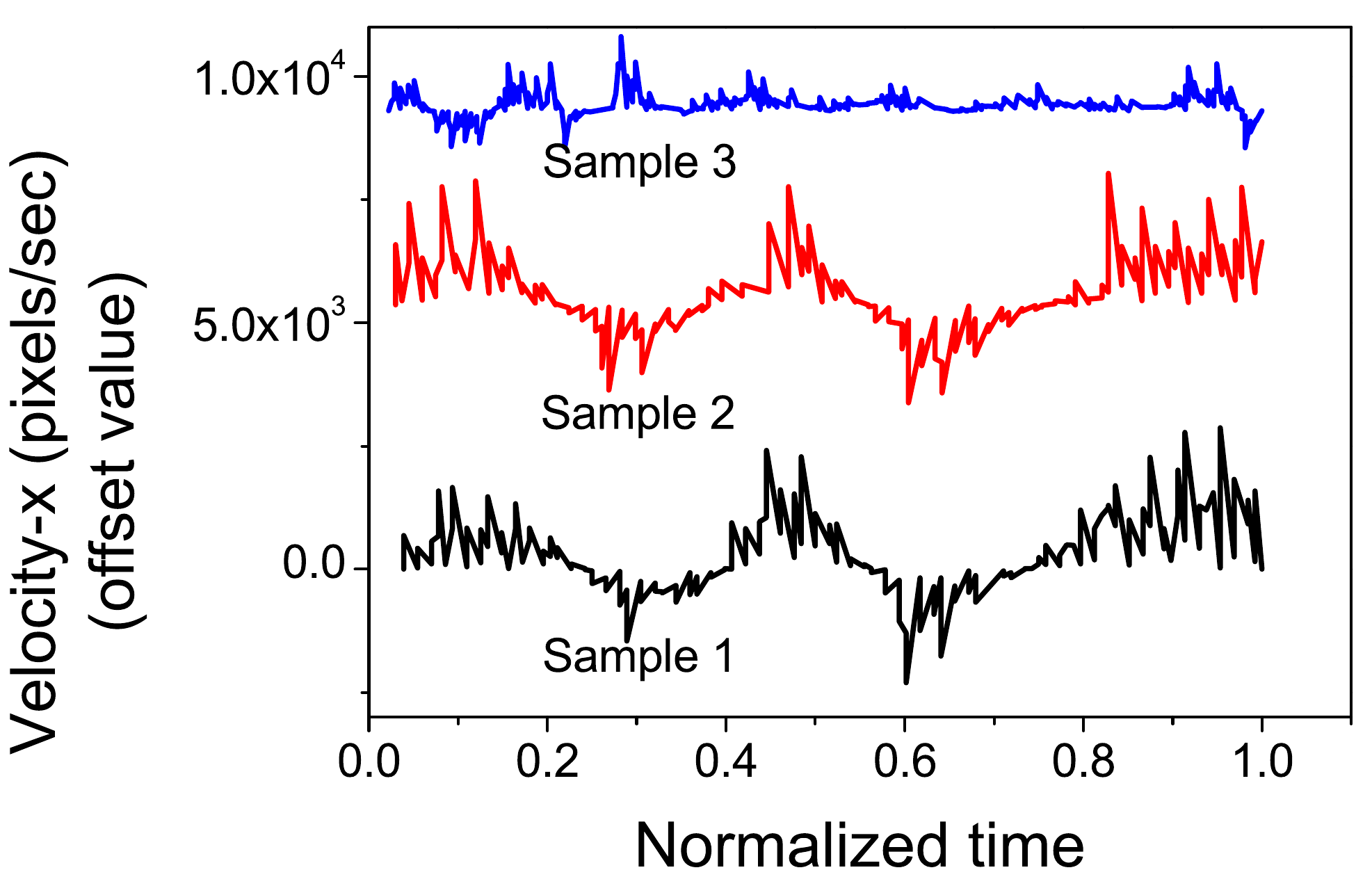}
    }
\hfill
    \subfigure[Velocity in y direction]{
        \label{yvelocity}
        \includegraphics[width=0.225\textwidth]{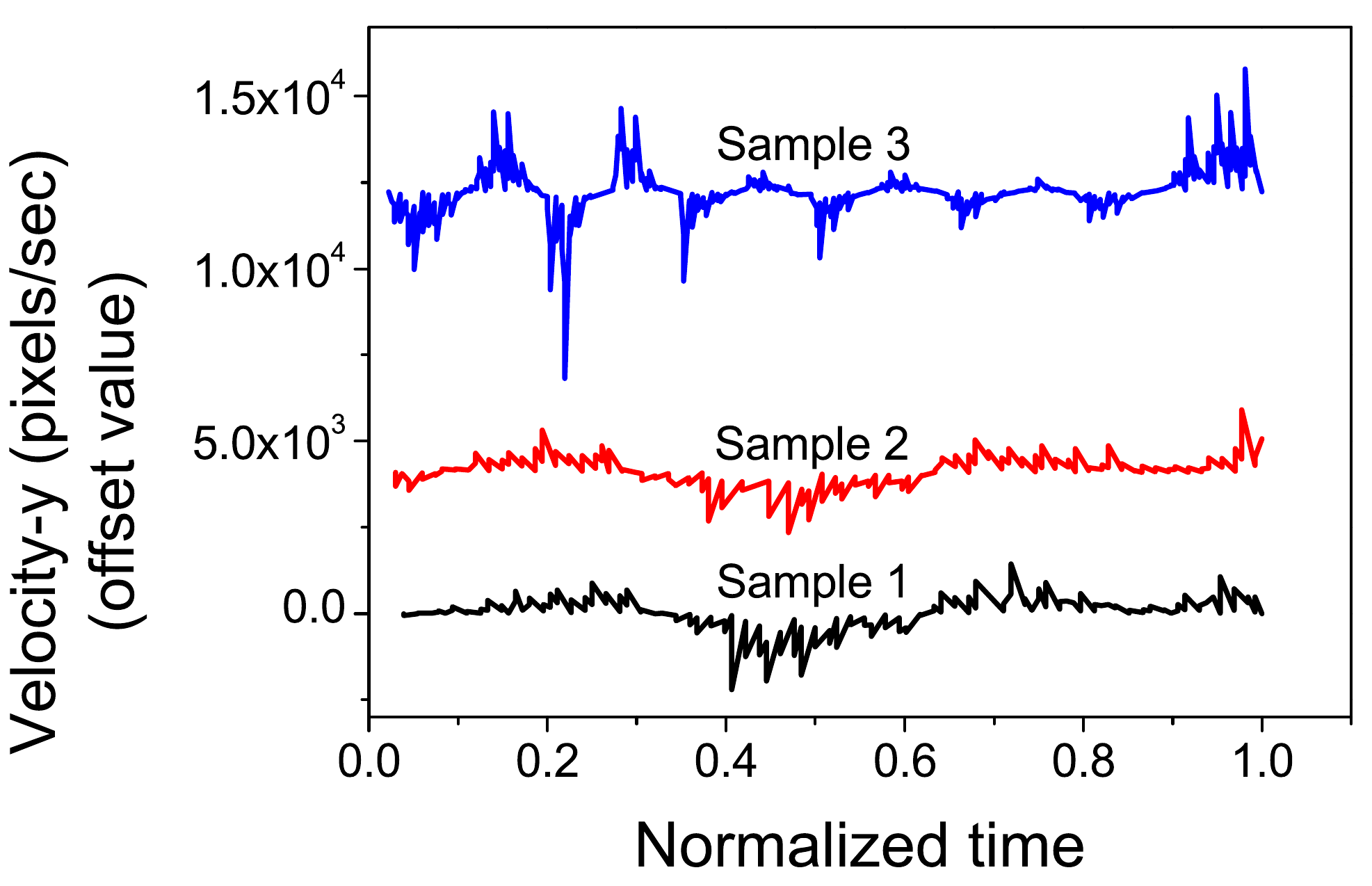}
    }
\caption{Velocity of curve drawing.}\label{velocity}
\end{figure}

\begin{figure}[!t]
\centering
    \subfigure[Acceleration in x direction]{
        \label{xacceleration}
        \includegraphics[width=0.225\textwidth]{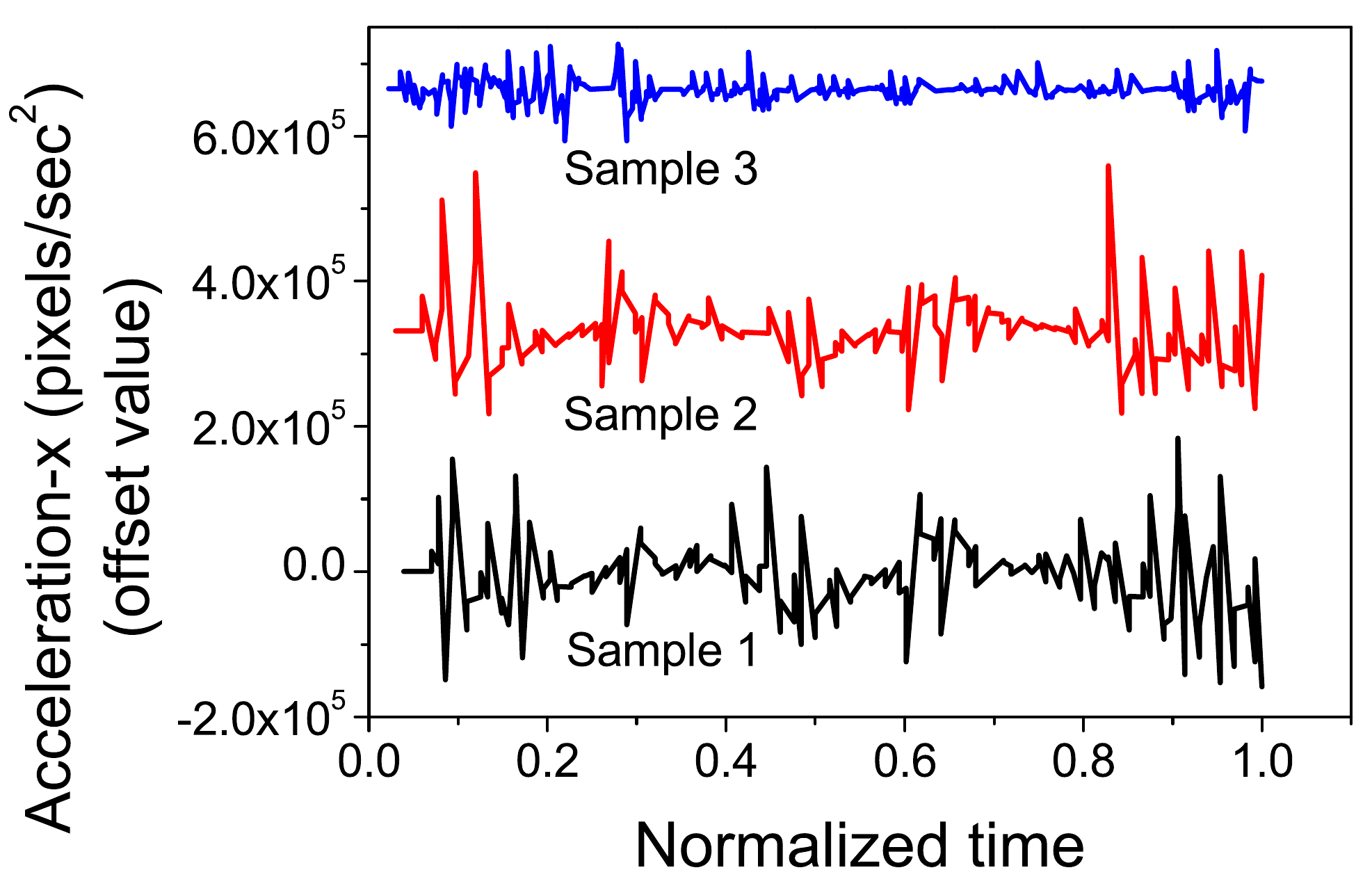}
    }
\hfill
    \subfigure[Acceleration in y direction]{
        \label{yacceleration}
        \includegraphics[width=0.225\textwidth]{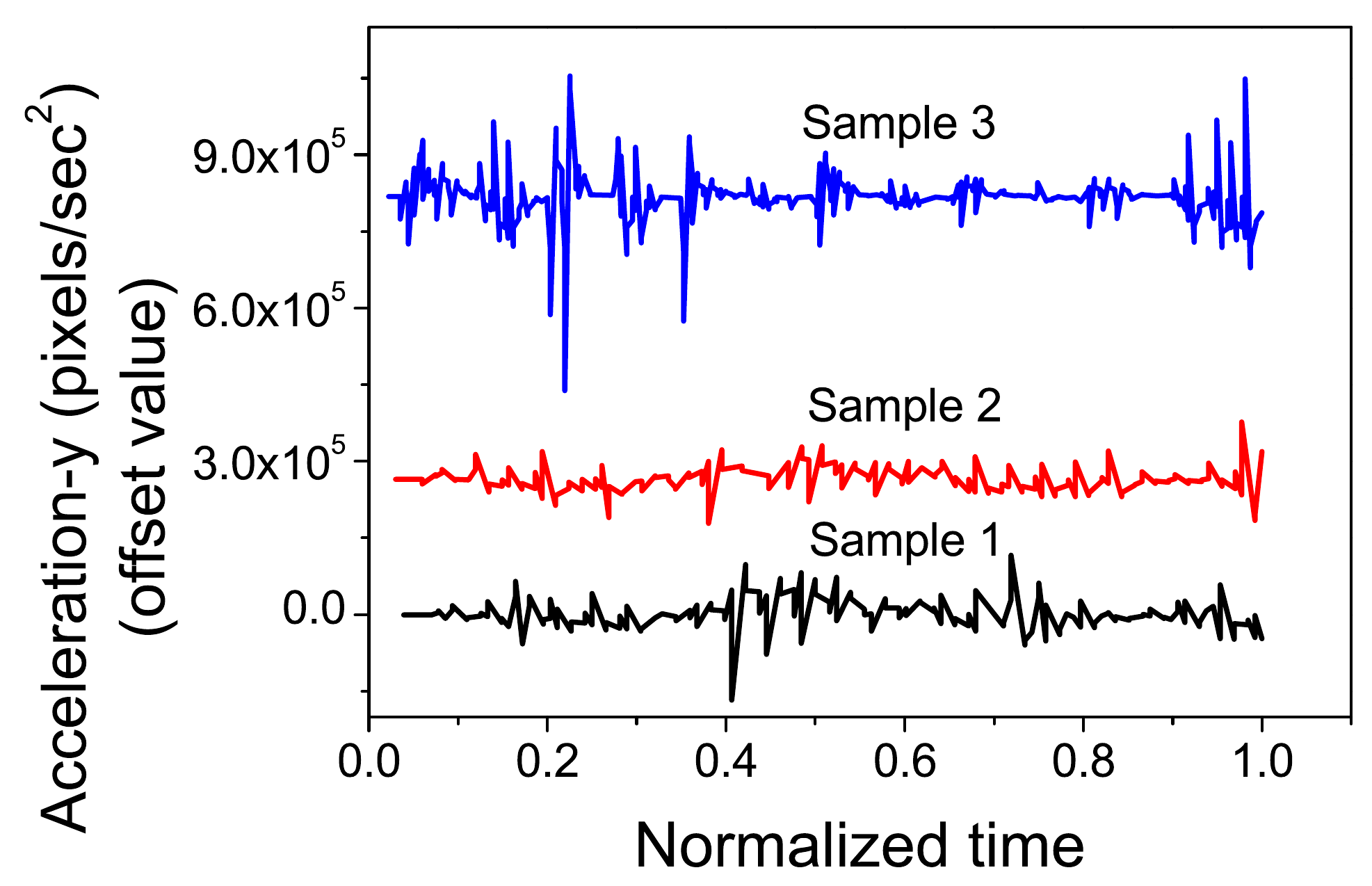}
    }
\caption{Acceleration of curve drawing.}\label{acceleration}
\end{figure}

\begin{figure}[!t]
\centering
    \subfigure[Curve direction]{
        \label{direction}
        \includegraphics[width=0.22\textwidth]{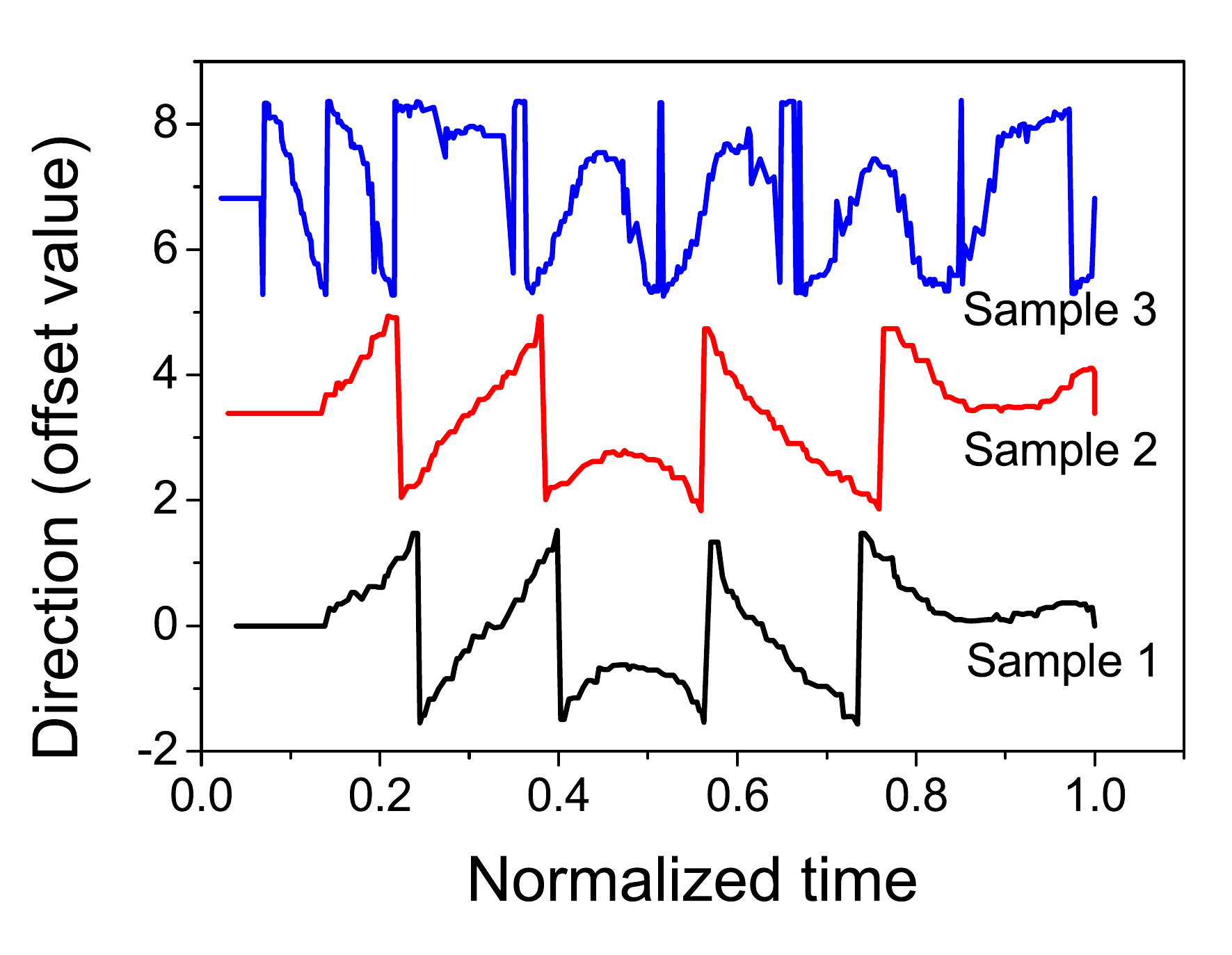}
    }
\hfill
    \subfigure[Curve curvature]{
        \label{curvature}
        \includegraphics[width=0.22\textwidth]{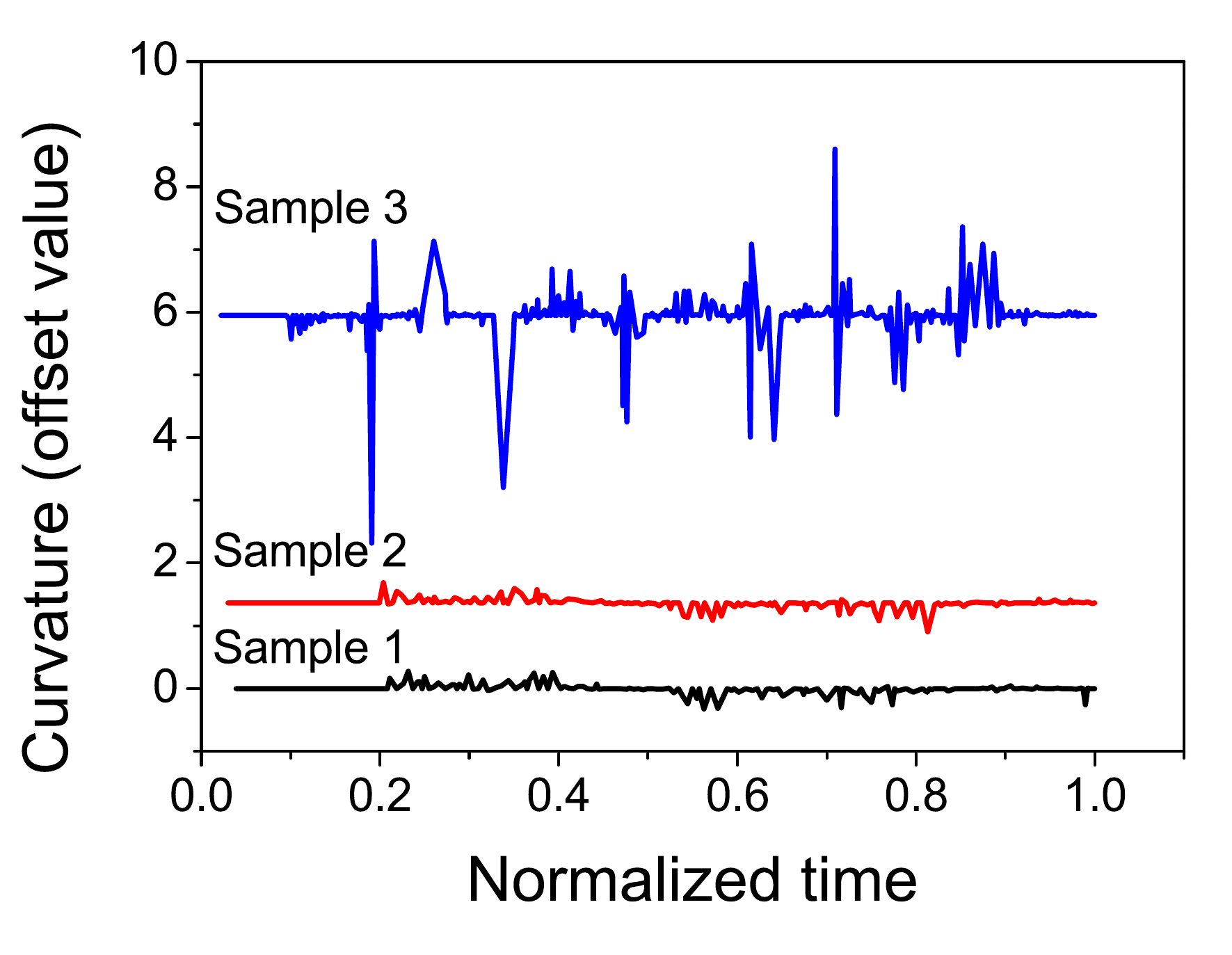}
    }
\caption{Direction and curvature.}\label{dirtcurv}
\end{figure}

\begin{figure}[!t]
\centering
    \subfigure[Touch pressure]{
        \label{pressure}
        \includegraphics[width=0.22\textwidth]{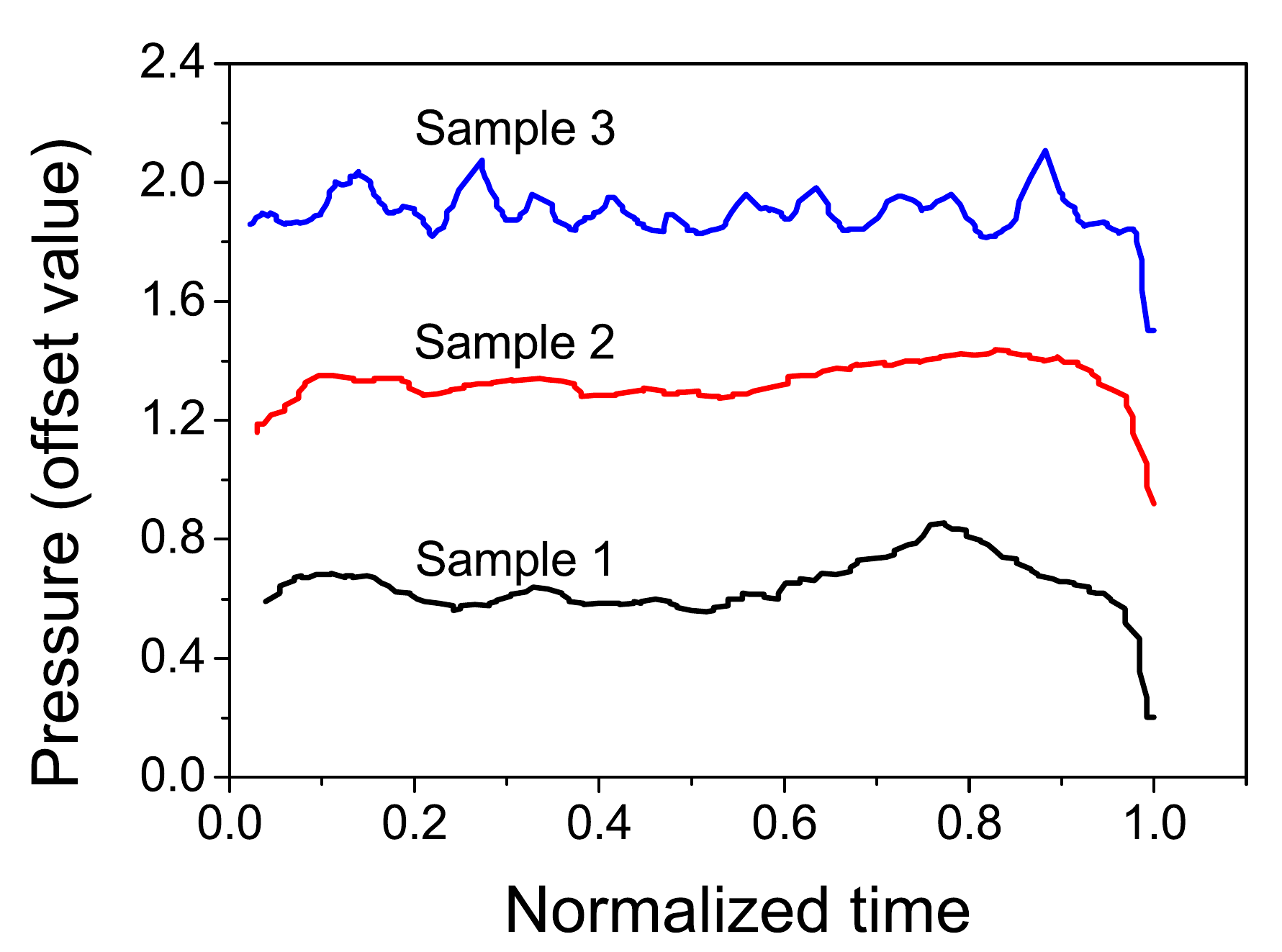}
    }
\hfill
    \subfigure[Touch size]{
        \label{size}
        \includegraphics[width=0.22\textwidth]{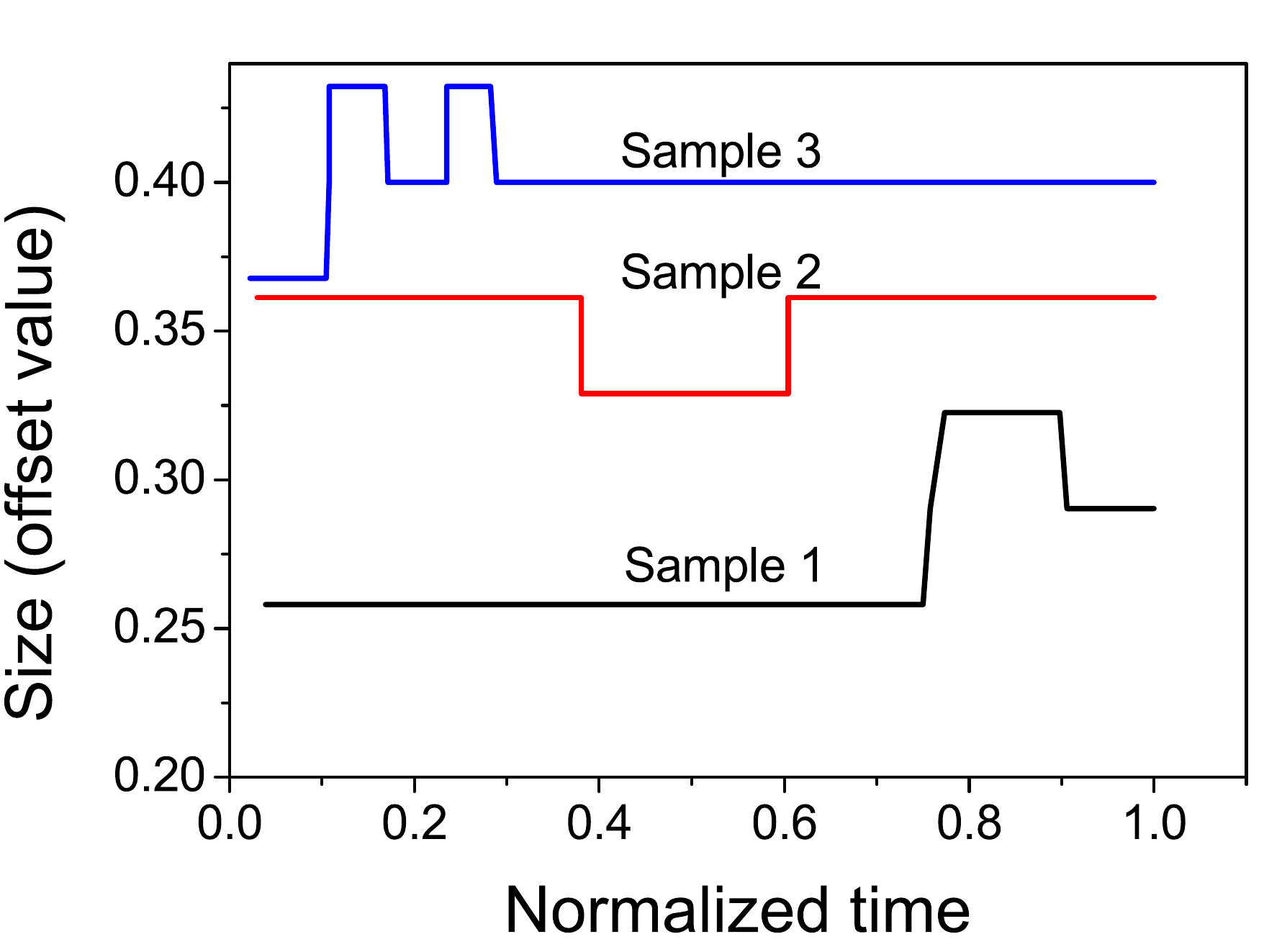}
    }
\caption{Touch pressure and size.}\label{pressize}
\end{figure}

\begin{figure}[!t]
\centering
    \subfigure[K-S test on single curve] {
        \label{KStest1}
        \includegraphics[width=3.5in]{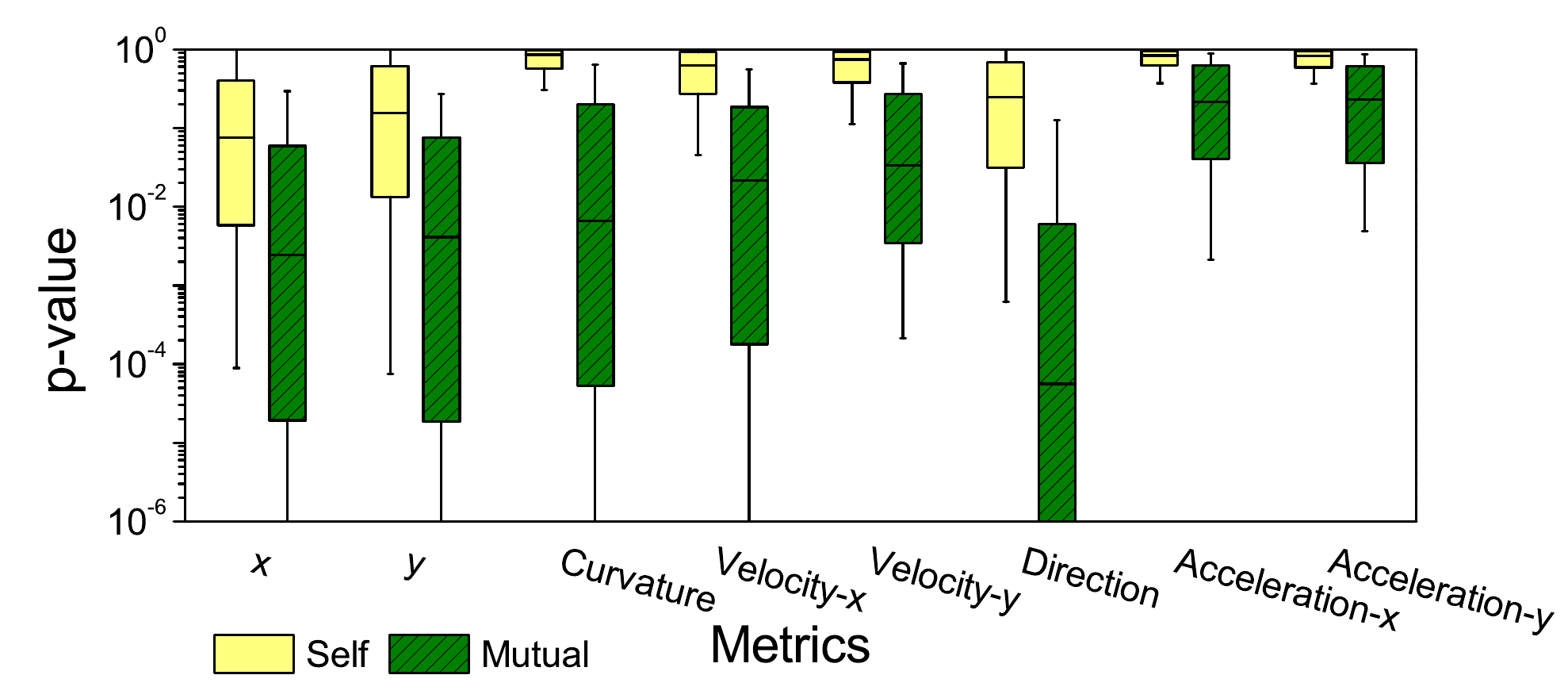}
    }
\hfill
    \subfigure[K-S test on multiple curves] {
        \label{KStest2}
        \includegraphics[width=3.5in]{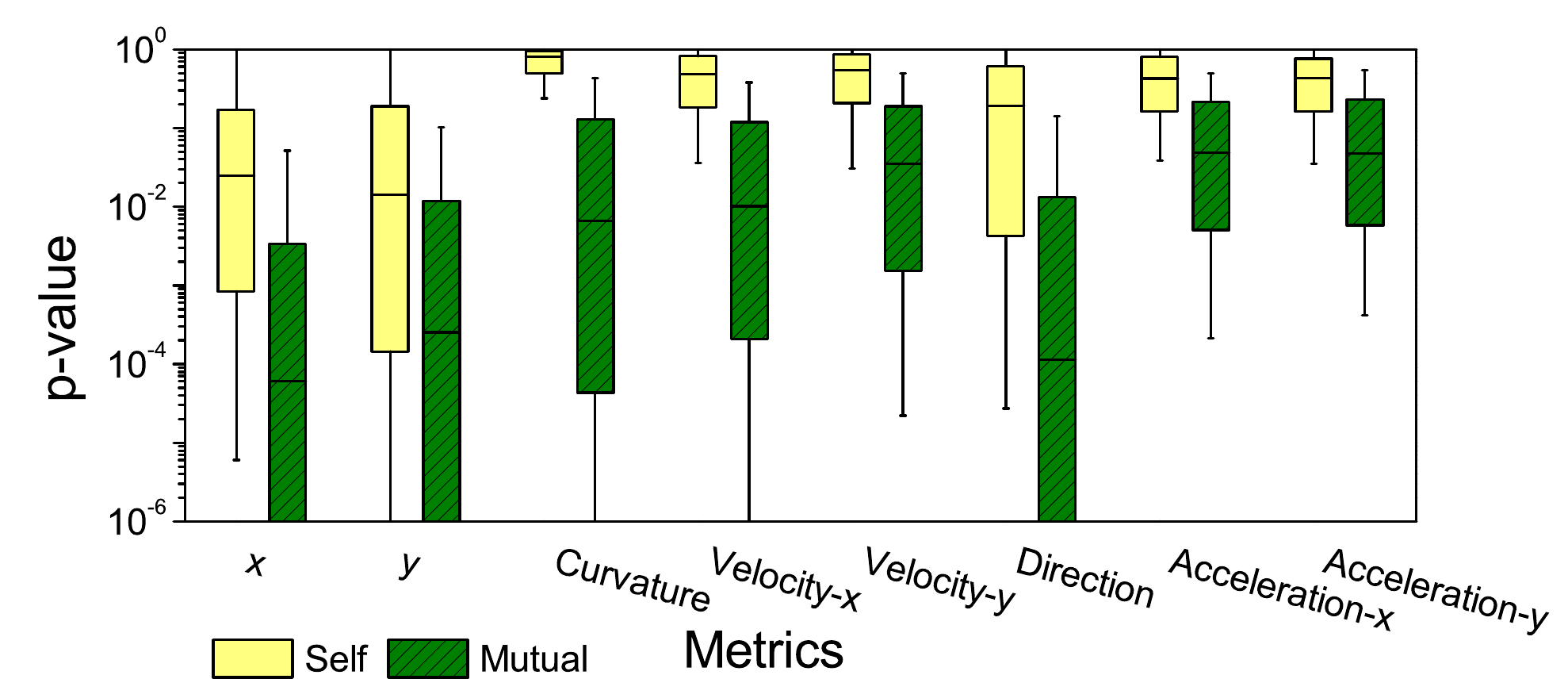}
    }
\hfill
\centering
\subfigure[K-S test on touch pressure and size] {
        \label{KStest3}
        \includegraphics[width=2.8in]{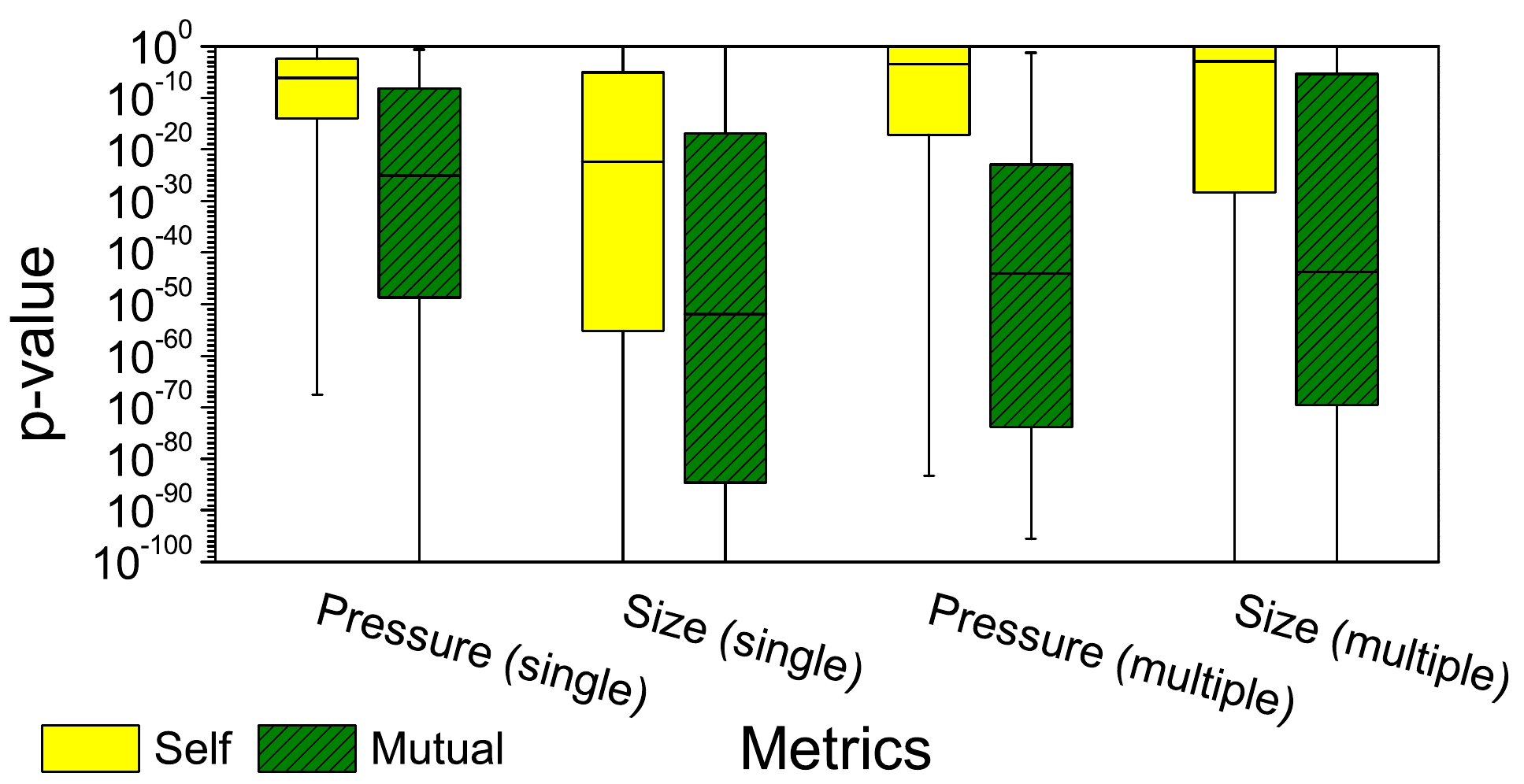}
    }
\caption{K-S test results.}\label{KStest}
\end{figure}

\subsection{Data Analysis}
We extract 10 time series from the touch event data associated with each single/multi-curve drawing, corresponding to 10 different curve features: x-coordinate, y-coordinate, curvature, x-velocity, y-velocity, x-acceleration, y-acceleration, direction, touch pressure, and touch size. The details of these features are further introduced in Section~\ref{subsec:FeatureSelection}.

Fig.~\ref{position} to Fig.~\ref{pressize} compare the curve features of two different samples of the same single-curve by the same randomly-chosen user (Samples 1 and 2) with the sample by a different randomly-chosen user (Sample 3). We can see that all the curve features above except touch size have quite consistent patterns for the same user and very different patterns for two different users. Similar results are obtained for multi-curves and omitted here for lack of space.

We further confirm the above observation via the two-sample Kolmogorov-Smirnov test (K-S test) on data collected from 10 users among 30 users to evaluate the statistic property of each curve feature. This leads to 780 single-curve samples and 780 multi-curve samples. The two-sample K-S test is the most useful and popular nonparametric method for comparing two samples. It outputs a $p$-value indicating the probability that two given samples are drawn from the same distribution or not. The larger the $p$-value is, the more likely the two samples come from the same distribution, and vice versa.

Fig.~\ref{KStest} shows the K-S test results, where the x-axis denotes different metrics, and the y-axis denotes the $p$-value. For each metric, the $p$-values are drawn in the box plot, and yellow empty (green shadowed) boxes show the distributions of $p$-values for different samples of the same user (of different users). The top and bottom of each box represent the 75th and the 25th percentiles of the $p$-value, respectively, and the middle bar denotes the median of the $p$-value. In addition, the upper and lower bars represent the upper and lower bounds of the $p$-value, respectively. For each metric, the less the overlap between the corresponding yellow empty box and green shadowed box, the better the metric is for distinguishing different users. We can observe that all the 10 curve features above except touch size are very good metrics for distinguishing different users, thus providing a solid basis for our TouchIn design.
\begin{figure*}[!t]
\centering
        \includegraphics[width=5.5in]{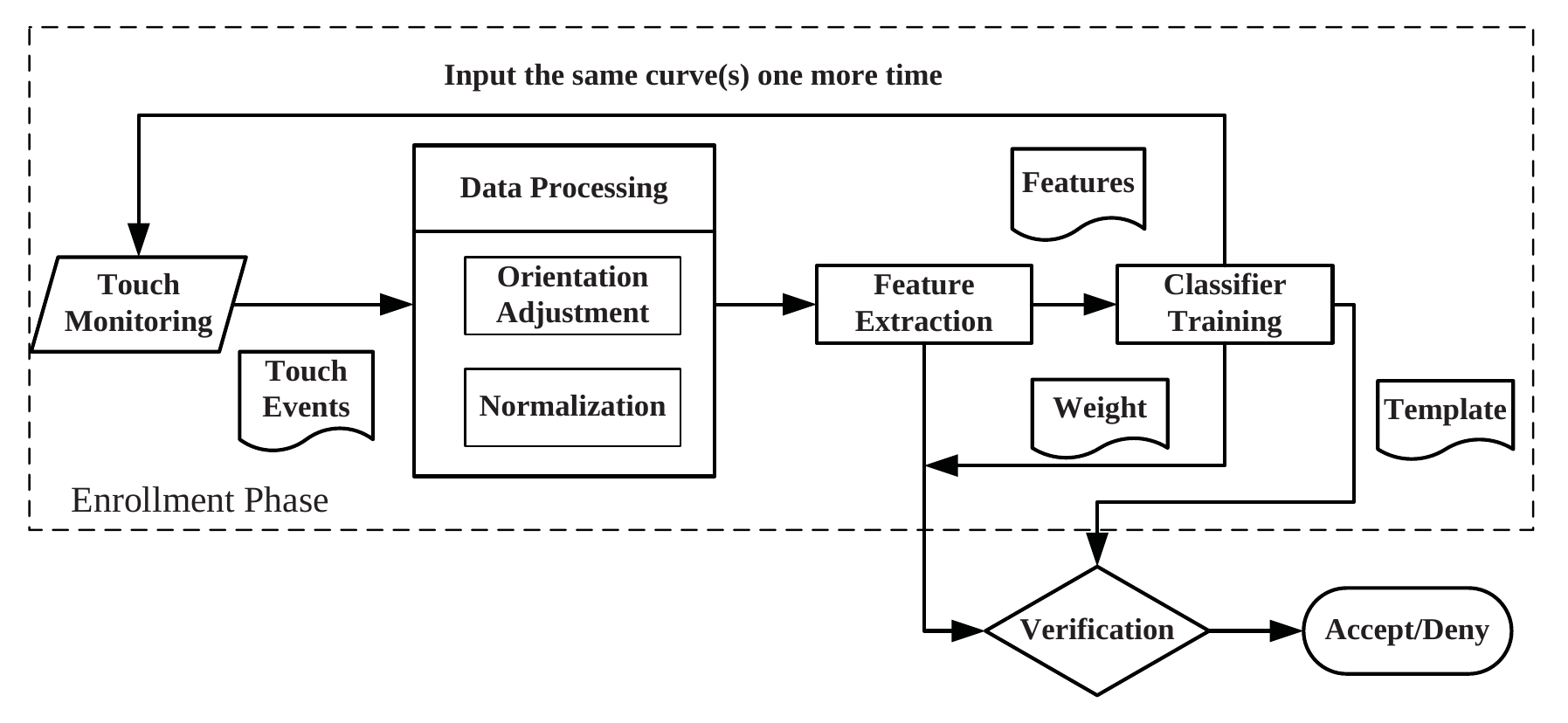}
\caption{The TouchIn system architecture.} \label{fig:systemarch}
\end{figure*}

\section{TouchIn Overview}\label{sec:DesignOverview}
In this section, we give an overview of the TouchIn system. As shown in Fig.~\ref{fig:systemarch}, TouchIn is composed of an enrollment phase and a verification phase.

The enrollment phase is to acquire an authentication template from the device owner's self-defined curves. In particular, the user is prompted to freely draw with one or multiple fingers of his choice on the touchscreen. A \texttt{Touch Monitoring }module is designed to track the motion of every finger by invoking Andriod APIs and record the touch-event data for every finger as a series of 4-tuples (finger ID, coordinate, timestamp, pressure). By connecting the coordinates in accordance with the timestamps, we can obtain the highly approximate curve drawn by every finger. TouchIn allows the user to hold the device in arbitrary ways and draw on any region of the touchscreen. This property is enabled by a \texttt{Data Processing} module to adjust the orientation of and normalize every curve. TouchIn uses ten features related to the geometric properties of a curve, how it is drawn, and who draws it, including x-position, y-position, direction, curvature, x-velocity, y-velocity, x-acceleration, y-acceleration, finger pressure, and hand geometry, respectively. So a \texttt{Feature Extraction} module is designed to extract hand geometry data, and a time series of feature values for each of the rest nine features. Finally, the extracted feature data are inputted into a \texttt{Classifier Training} module to generate an authentication template, which is stored on the mobile device. The \texttt{Classifier Training} module also outputs the weights assigned to every feature when composing the authentication template. Generating a high-quality template may need the device owner to draw multiple times in (approximately) the same way, which we will seek to minimize for usability concerns.

During the verification phase, anyone attempting to unlock the device is prompted to input a password without any hint given. Every finger's input will be treated as a single curve. The same first three modules in the enrollment phase are used to extract ten features from every input. Then the feature data from all finger inputs are input into a \texttt{Verification} module for comparison with the authentication template. If there is a match, the user is considered legitimate and allowed in; otherwise, the user can retry. The user is considered unauthorized and denied access after a threshold number of failed attempts.

\section{Enrollment}\label{Sec:Enrollment}
In this section, we detail the design of the enrollment phase in accordance with the diagram in Fig.~\ref{fig:systemarch}.

\subsection{Data Processing}
This module is invoked after the device owner finishes drawing by lifting his fingers and takes the resulting 4-tuple touch-event data as input. The following submodules are then executed sequentially.

\subsubsection{Orientation Adjustment}
TouchIn does not require the user to hold the device in any specific way or draw in designated regions. As said, this user-friendly feature is important for satisfying the sightless requirement as well as thwarting shoulder-surfing and smudge attacks. The consequence is that multiple attempted drawings of the same curve on different touchscreen regions will lead to different sets of touch coordinates, which are unlikely to compare. To handle this situation, we design two methods to adjust the curve orientations, which apply when one or multiple curves are detected, respectively.

\vspace{.05in}\noindent \textbf{\underline{Single-curve orientation adjustment}}

\begin{figure}[t]
\centering
        \includegraphics[width=2.7in]{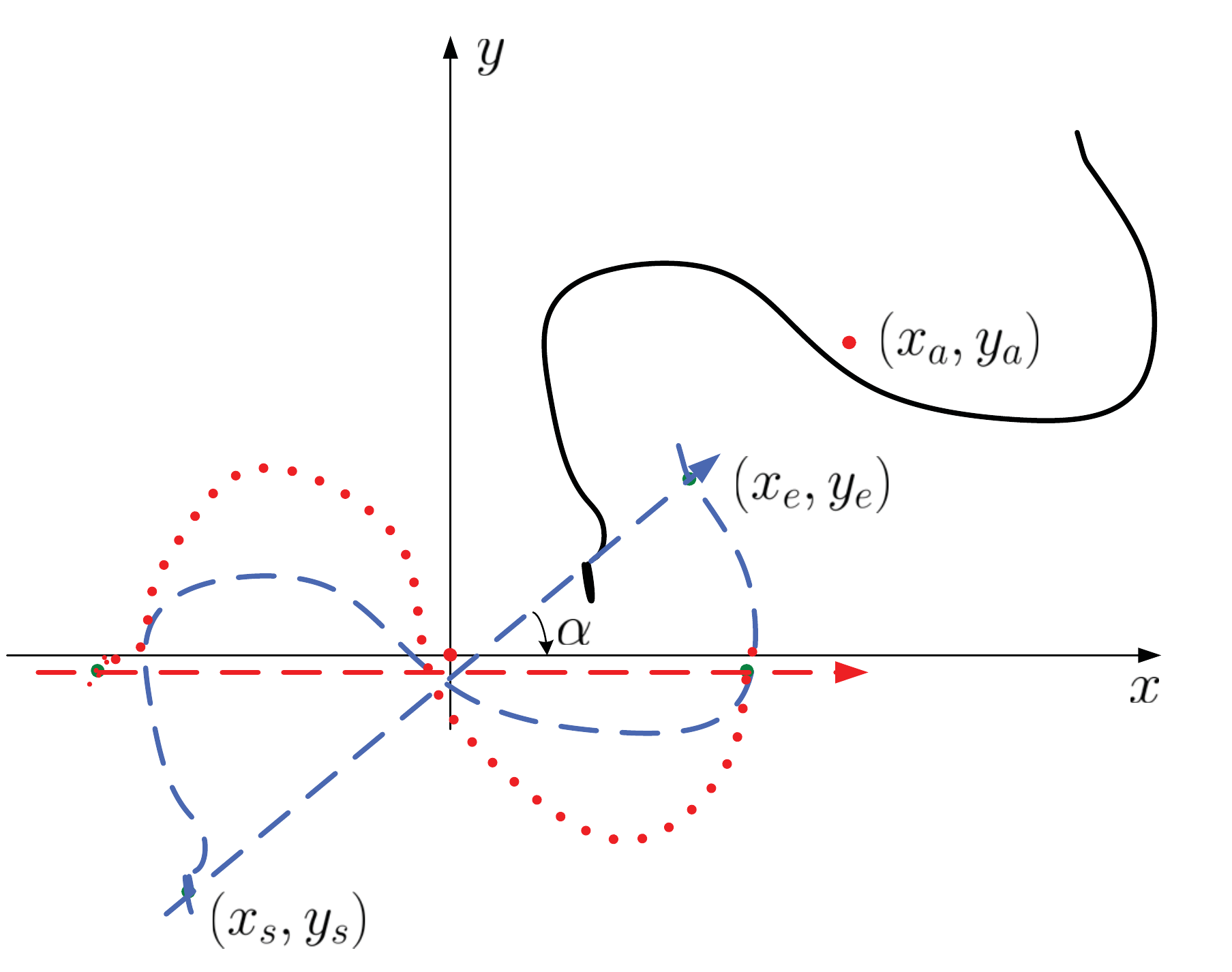}
\caption{Single-curve orientation adjustment: case 1.} \label{fig:orien1}
\end{figure}

\vspace{.1in}The basic idea of this technique is to adjust the curve orientation based on some feature points of the curve. Assume that the curve is associated with $l$ touch coordinates denoted by $\{(x_i, y_i)\}_{i=1}^{l}$. We first compute the coordinate $(x_a,y_a)$ of the curve's center point as $x_{a} = \sum_{i=1}^{l}x_{i}/l$ and $y_{a} = \sum_{i=1}^{l}y_{i}/l$. Then we move the curve along horizontal and vertical directions until the center point becomes the origin of a Cartesian coordinate system. After this movement, the $l$ coordinates become $\{(x'_i, y'_i)\}_{i=1}^{l}$, where $x'_{i} = x_{i}-x_{a}$, and $y'_{i} =y_{i}-y_{a}$. The next step is to decide the starting and ending points of the curve. Note that every finger contacts with the touchscreen will generate some closer coordinates with almost identical timestamps, and it is unlikely to reproduce every finger contact. This means that even if we want to draw the same curve in exactly the same way, the starting and ending points of the curve may still be different. To accommodate this variation, we compute an average starting point with coordinate $(x_{s}, y_{s})$ as the average of the first $\eta_s$ coordinates in time and an average ending point with coordinate $(x_{e}, y_{e})$ as the average of the last $\eta_e$ coordinates in time, where $\eta_s$ and $\eta_e$ are empirical system parameters. Finally, we draw an arrow from $(x_{s}, y_{s})$ to $(x_{e}, y_{e})$ and then rotate the curve until the arrow direction is the same as that of the $x$-axis. After this rotation, the $l$ coordinates of the curve become $\{(x''_i, y''_i)\}_{i=1}^{l}$, where $x''_{i} = x'_{i}\cos\alpha + y'_{i}\sin\alpha$, $y''_{i} = -x'_{i}\sin\alpha + y'_{i}\cos\alpha,$ and $\alpha = \arctan\frac{y_{e}-y_{s}}{x_{e}-x_{s}}$. Fig.~\ref{fig:orien1} shows an example, where the solid, dashed, and dotted curves correspond to the original one, the one after movement, and the one after rotation, respectively.

\begin{figure}[t]
\centering
        \includegraphics[width=2.7in]{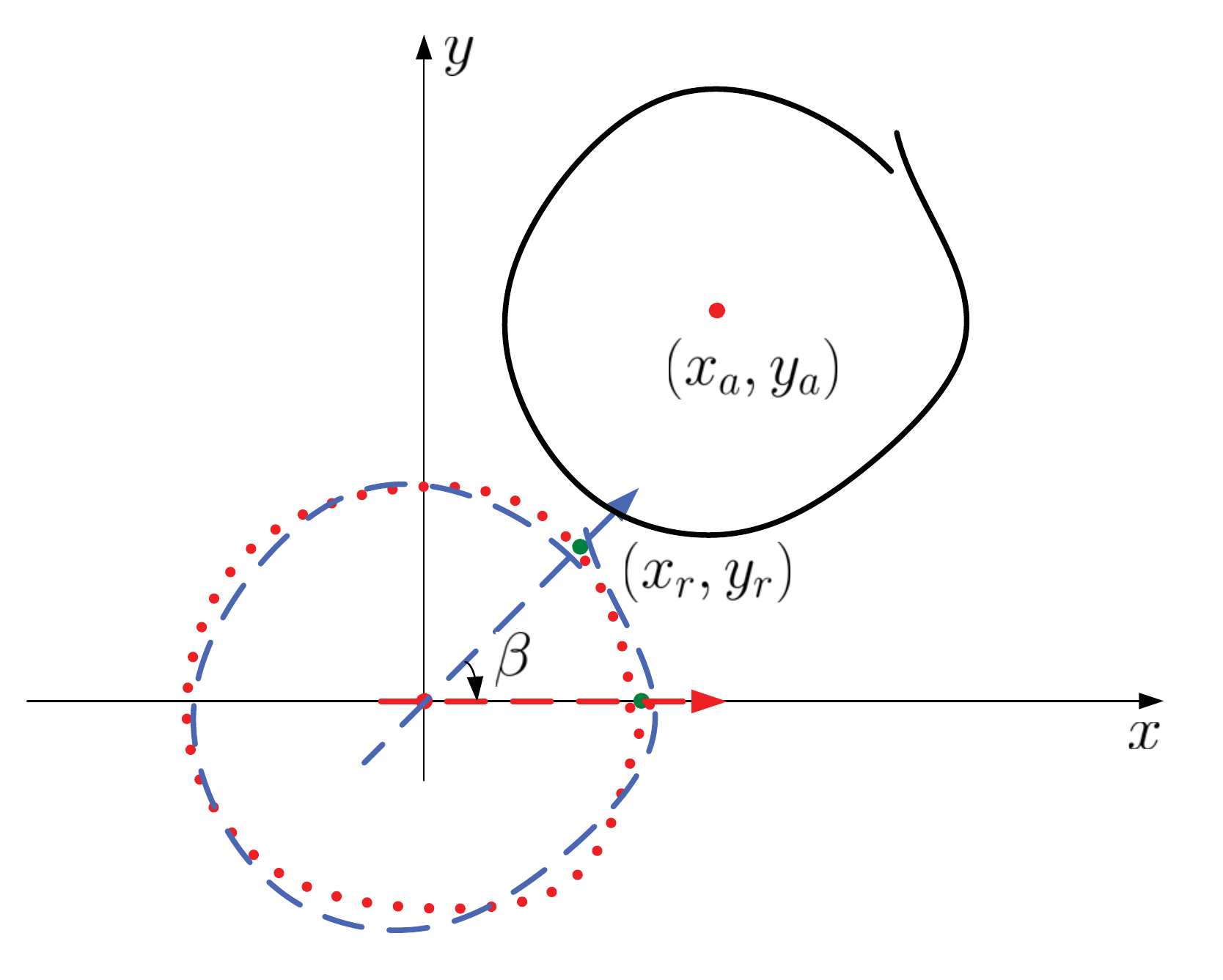}
\caption{Single-curve orientation adjustment: case 2.} \label{fig:orien2}
\end{figure}

If the distance between the average starting and ending points is very small, i.e., $\sqrt{(x_{s} - x_{e})^2 + (y_{s} - y_{e})^2} \leq \xi$ for a very small system parameter $\xi$, the average starting and ending points are likely to be different even if the same user tries to draw the same curve multiple times. As a result, the very short arrow from the average starting point to the average ending point may have opposite directions for different drawings of the same curve. If we still rotate the curve as above, the same user's multiple drawings may not match. To handle this situation, we further define an anchor point with coordinate $(x_r, y_r)$, where $x_r = (x_{s} + x_{e})/2$, and $y_r = (y_{s} + y_{e})/2$. Then we draw an arrow from the origin to the anchor point and rotate the curve until the arrow direction is the same as that of the $x$-axis. Finally, the $l$ coordinates of the curve become $\{(x''_i, y''_i)\}_{i=1}^{l}$, where $x''_{i} = x'_{i}\cos\beta + y'_{i}\sin\beta$, $y''_{i} = -x'_{i}\sin\beta + y'_{i}\cos\beta,$ and $\beta= \arctan\frac{y_{r}}{x_{r}}$. This technique, however, cannot be applied when the average starting and ending points are sufficiently far from each other. An example for this case is shown in Fig.~\ref{fig:orien2}, where the solid, dashed, and dotted curves correspond to the original one, the one after movement, and the one after rotation, respectively.

\vspace{.2in}\noindent \textbf{\underline{Multi-curve orientation adjustment}}

\begin{figure}[t]
\centering
        \includegraphics[width=2.7in]{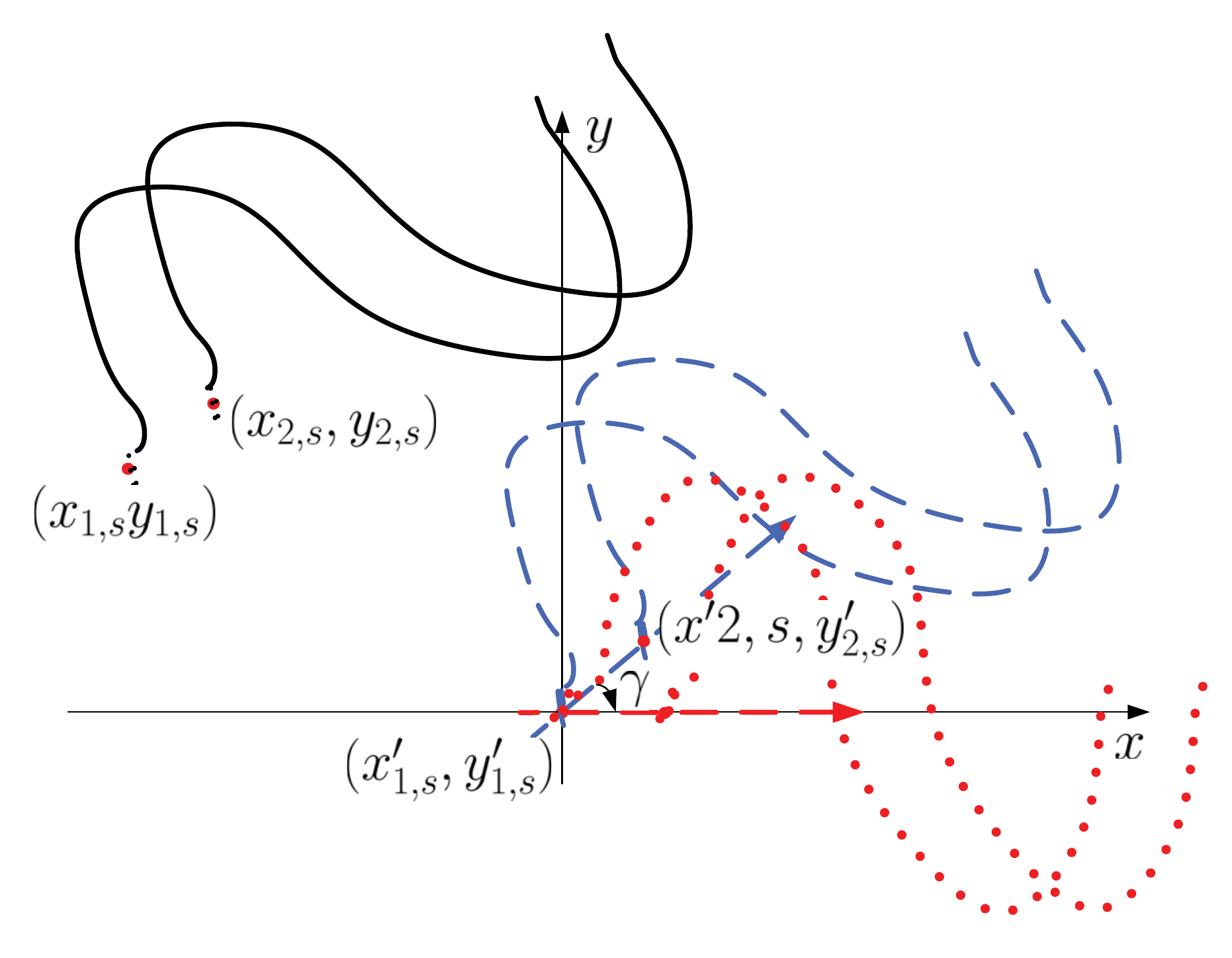}
\caption{Multi-curve orientation adjustment.} \label{fig:orien3}
\end{figure}

\vspace{.1in} The orientations of multiple curves can be similarly adjusted. In particular, assume that the device owner uses $M\in [2,5]$ fingers to simultaneously draw $M$ curves. We denote the original coordinates of the $i$th ($\forall i \in[1,M]$) curve by $\{(x_{i,j}, y_{i,j})\}_{j=1}^{l_i}$. We first compute the average starting point of the first curve with coordinate $(x_{1,s},y_{1,s})$ as the average of the first $\eta_s$ ones in $\{(x_{1,j}, y_{1,j})\}_{j=1}^{l_1}$. Then we move the $M$ curves together along horizontal and vertical directions until the average starting point becomes the origin. The $l_i$ coordinates of the $i$th ($\forall i \in[1,M]$) curve thus become $\{(x'_{i,j}, y'_{i,j})\}_{j=1}^{l_i}$, where $x'_{i,j} = x_{i,j} - x_{1,s}$, and $y'_{i,j} = y_{i,j} - y_{1,s}$. Then we compute the average starting point of the second curve with coordinate $(x'_{2,s},y'_{2,s})$ as the average of the first $\eta_s$ ones in $\{(x'_{2,j}, y'_{2,j})\}_{j=1}^{l_2}$. Finally, we draw an arrow from the origin to $(x'_{2,s},y'_{2,s})$ and rotate the $M$ curves together until the arrow direction follows the $x$-axis. After this rotation, the $l_i$ coordinates of the $i$th ($\forall i \in[1,M]$) curve change to $\{(x''_{i,j}, y''_{i,j})\}_{j=1}^{l_i}$, where
$x''_{i,j} = x'_{i,j}\cos\gamma + y'_{i,j}\sin\gamma,$ $y''_{i,j} = -x'_{i,j}\sin\gamma + y'_{i,j}\cos\gamma,$ and $\gamma = \arctan\frac{y'_{2,s}}{x'_{2,s}}$. Fig.~\ref{fig:orien3} gives an example where the solid, dashed, and dotted curves refer to the original ones, the ones after movement, and the ones after rotation, respectively.

For this method to work, the device owner needs to put his $M$ fingers always in the same order on the touchscreen, which is fairly easy according to our experimental studies. This requirement also implies that the touching order of his $M$ fingers can also be implicitly used as a partial password. Specifically, although the adversary may manage to observe the shapes of the $M$ curves and reproduce them, the order in which the $M$ fingers touch the screen is much more subtle to observe. As a result, the $M$ curves drawn by the adversary may be ordered very differently from those drawn by the device owner. It follows that the adversary's $M$ curves after the above orientation adjustment may be very different from those of the device owner, in which case the authentication attempt will fail.

\subsubsection{Normalization}
Even if the device owner can accurately reproduce his curve shapes, it is often more difficult for him to memorize and reproduce the curve lengths. The normalization submodule is designed to handle this situation and applies to both single-curve and multi-curve cases. Consider the multi-curve case as an example. Recall that the $l_i$ coordinates of the $i$th curve change to $\{(x''_{i,j}, y''_{i,j})\}_{j=1}^{l_i}$. We define the normalized $l_i$ coordinates as $\{(\bar{x}_{i,j}, \bar{y}_{i,j})\}_{j=1}^{l_i}$, where
\[
\bar{x}_{i,j} = \frac{x''_{i,j}-\min\{x''_{i,k}\}_{k=1}^{l_i}}{\max\{x''_{i,k}\}_{k=1}^{l_i}-\min\{x''_{i,k}\}_{k=1}^{l_i}}\;,
\]
and
\[
\bar{y}_{i,j} = \frac{y''_{i,j}-\min\{y''_{i,k}\}_{k=1}^{l_i}}{\max\{y''_{i,k}\}_{k=1}^{l_i}-\min\{y''_{i,k}\}_{k=1}^{l_i}}\;.
\]
The same method also applies to the single-curve case.

\subsection{Feature Extraction}\label{subsec:FeatureSelection}
Then we extract ten features from the touch event data for every curve. The following description applies to the $i$th curve ($\forall i\in [1,M]$) of the multi-curve case and also applies to the single-curve case after slightly changing the notation.
\begin{itemize}
    \item \textbf{x- and y-coordinates}: $\mathcal{C}_i=\{(\bar{x}_{i,j}, \bar{y}_{i,j})\}_{j=1}^{l_i}$.

    \item \textbf{Curvature}:  The curvature is the amount by which a geometric object deviates from being flat or straight. The curvature at point $(\bar{x}_{i,j}, \bar{y}_{i,j})$ can be derived as
        \begin{equation}
        \kappa_{i,j} = \frac{4\Psi^y_{i,j}\Delta^x_{i,j} - 4\Psi^x_{i,j}\Delta^y_{i,j}}{((\Delta^x_{i,j})^2+(\Delta^y_{i,j})^2)^{3/2}}\;,
        \end{equation}
         where
         \[
         \begin{split}
         \Delta^x_{i,j}&= (\bar{x}_{i,j+1}-\bar{x}_{i,j-1})/2,\\
         \Delta^y_{i,j}&= (\bar{y}_{i,j+1}-\bar{y}_{i,j-1})/2,\\
         \Psi^y_{i,j}& = (\bar{y}_{i,j+1}-2\bar{y}_{i,j}+\bar{y}_{i,j-1}),\\
         \Psi^x_{i,j}& = (\bar{x}_{i,j+1}-2\bar{x}_{i,j}+\bar{x}_{i,j-1})\;.
         \end{split}
         \]

    \item \textbf{x- and y-velocity}: The x-velocity and y-velocity of touch-event $j\in[2,l_i]$ are defined as
        \begin{equation}
        \mathcal{V}_{i,j}^x = \frac{\bar{x}_{i,j}-\bar{x}_{i,j-1}}{t_{i,j}-t_{i,j-1}},
        \end{equation}
        \begin{equation}
        \mathcal{V}_{i,j}^y = \frac{\bar{y}_{i,j}-\bar{y}_{i,j-1}}{t_{i,j}-t_{i,j-1}},
        \end{equation}
        where $t_{i,j}$ and $t_{i,j-1}$ denote the timestamp of the $j$th and $(j-1)$th touch event, respectively.

    \item \textbf{x- and y-acceleration}: The acceleration of the $j$th touch event is defined as
        \begin{equation}
        \mathcal{A}_{i,j}^x = \frac{\mathcal{V}_{i,j}^x-\mathcal{V}_{i,j-1}^x}{t_{i,j}-t_{i,j-1}},
        \end{equation}
        \begin{equation}
        \mathcal{A}_{i,j}^y = \frac{\mathcal{V}_{i,j}^y-\mathcal{V}_{i,j-1}^y}{t_{i,j}-t_{i,j-1}}.
        \end{equation}

    \item \textbf{Direction}: The direction at $(\bar{x}_{i,j}, \bar{y}_{i,j})$ is defined as
        \begin{equation}
        \alpha_{i,j} = \arctan\frac{\bar{y}_{i,j+1}-\bar{y}_{i,j}}{\bar{x}_{i,j+1}-\bar{x}_{i,j}}.
        \end{equation}

    \item \textbf{Touch pressure}: The touch pressure on the touchscreen is hardly observable by the adversary and can thus also be used as some behavioral characteristic. Android OS reports the finger pressure for every touch event as a value ranging from 0 (no pressure at all) to 1 (normal pressure), and can be denoted as $\mathcal{P}_i=\{p_{i,j}\}_{j=1}^{l_i}$.

    \item \textbf{Hand geometry}: Due to the uniqueness of each user's hand size and shape, the distance between any two fingers (known as hand geometry information) can also be used to uniquely identify a user. $M$ fingers together determine $\binom{M}{2}$ finger pairs, where the distance between the $i$th and $j$th fingers is computed as
        \begin{equation}
        \ell_{i,j} = \sqrt{(\bar{x}_{i,s} - \bar{x}_{j,s})^2 + (\bar{y}_{i,s} - \bar{y}_{j,s})^2}
        \end{equation}
        where $(\bar{x}_{i,s}, \bar{y}_{i,s})$ and $(\bar{x}_{j,s}, \bar{y}_{j,s})$ denote the normalized coordinates of the average starting points of the $i$th and $j$th curves, respectively. These $\binom{M}{2}$ distance metrics compose the hand geometry information.
\end{itemize}

\subsection{Classifier Training}
This module is to obtain an optimal classifier based on the extracted feature data from the training set. In what follows, we first discuss the comparison of two arbitrary curves. We then discuss how to assign optimal weights to coordinate, curvature, velocity, acceleration, and direction features. Since hand geometry information does not have an associated time series, we discuss how to incorporate it into the authentication template in Section~\ref{sec:Selection}.

\subsubsection{Curve Comparison}\label{SubsubSec:DistCal}
As said, every finger-drawn curve is associated with nine time series, corresponding to the x-coordinate, y-coordinate, curvature, x-velocity, y-velocity, x-acceleration, y-acceleration, direction and touch pressure features, respectively. To compare two arbitrary curves, we use Dynamic Time Warping (DTW) \cite{SakoeDyn78} to compute the distance between each of the nine time-series pairs. For example, consider one pair of time series, denoted by $S_1$ and $S_2$ with $n_1$ and $n_2$ feature values, respectively, where $n_1$ may not equal $n_2$. To compare $S_1$ and $S_2$, we first construct an $n_1\times n_2$ distance matrix, in which the value in each cell $(i, j)$ represents the absolute distance between the $i$th feature value of $S_1$ and the $j$th feature value of $S_2$. DTW finds the shortest path from cell $(1, 1)$ to cell $(n_1, n_2)$, along which the total value of the cells is defined as the distance between $S_1$ and $S_2$. We denote by $\{d_i\}_{i=1}^9$ the DTW distance results for the x-coordinate, y-coordinate, curvature, x-velocity, y-velocity, x-acceleration, y-acceleration, direction and touch pressure features, respectively. Then we need to find a combination of $\{d_i\}_{i=1}^9$ to classify the curves from both authorized and unauthorized users, where $w_i$ denotes the weight assigned to feature $i$.

\subsubsection{Weight Assignment}\label{sec:Weight}
The objective of weight assignment is to assign different weights to different features to minimize false negatives and positives. For this purpose, TouchIn will come with $\lambda$ random curves, each also having nine time series of feature data. Assume that the device owner is prompted to draw every chosen curve $\omega$ times, where each drawing is referred to as a sample curve. These $\omega$ sample curves along with the $\lambda$ preloaded curves compose a training set. Then we randomly select one of the $\omega$ sample curves as a reference curve and use DTW to compare it with every other curve in the training set. For convenience, we denote the reference curve by CURVE$_0$ and every other curve by CURVE$_i$ for $i\in [1,\omega+\lambda-1]$. We also let $\{d_{i,j}\}_{j=1}^9$ denote the nine DTW distance results between CURVE$_0$ and CURVE$_i$.

We formulate weight assignment as a classification problem. In particular, all the training curves can be classified into two classes: the $\omega$ sample curves of the device owner belong to Class I ($\ell=0$), while the $\lambda$ random curves belong to Class II ($\ell=1$). We use the Logistic Regression (LR) algorithm \cite{AlpayInt10} as the classification algorithm to distinguish the two classes. In particular, for a two-classification task, the LR algorithm finds the optimal linear combination of all features to separate two classes of curves with the minimum misclassification cost. For this purpose, we further introduce a constant $d_0=1$ and $w_0$ as the corresponding weight for $d_0$. Let $\mathbf{d}_i = [d_0, d_{i,1}, d_{i,2}, \cdots, d_{i,9}]^T$ be the distance vector and $\mathbf{w} = [w_0, w_1, w_2, \cdots, w_9]^T$ be the weight vector of CURVE$_i$ ($\forall i\in [1,\omega+\lambda-1]$). The linear combination of the distance and weight vectors is expressed by $\mathbf{w}^T\mathbf{d}_i = w_0 + w_1d_{i,1} + w_2d_{i,2} + \cdots + w_9d_{i,9}$. We proceed to define
\begin{equation}
h(\mathbf{w}^T\mathbf{d}_i)=\frac{1}{1+e^{-\mathbf{w}^T\mathbf{d}_i}},
\end{equation}
where $h(\cdot)$ is the sigmoid function with the range $[0,1]$. In particular, if $\mathbf{w}^T\mathbf{d}<0$, then $h(\mathbf{w}^T\mathbf{d})< 0.5$, and if $\mathbf{w}^T\mathbf{d}>0$, then $h(\mathbf{w}^T\mathbf{d})\geq 0.5$. Then CURVE$_i$ can be classified as
\begin{align}
\ell_i = \begin{cases}
      0, & h(\mathbf{w}^T\mathbf{d}_i) < 0.5, \\
      1, & h(\mathbf{w}^T\mathbf{d}_i) \geq 0.5\;.
    \end{cases}
\end{align}
We need to make sure that the majority of the sample curves can be classified into Class I (low false negatives) and most of the $\lambda$ preloaded curves can be classified into class II (low false positives). An ideal classifier should have $h(\mathbf{w}^T\mathbf{d}_i)$ much smaller (or larger) than 0.5 if CURVE$_i$ is a sample (or preloaded) curve. We thus define the misclassification cost for CURVE$_i$ as
\begin{align}
C(h(\mathbf{w}^T\mathbf{d}_i),\ell_i) = \begin{cases}
      -\log(1-h(\mathbf{w}^T\mathbf{d}_i)), & \ell_i = 0, \\
      -\log(h(\mathbf{w}^T\mathbf{d}_i)), & \ell_i = 1.
    \end{cases}
\end{align}
The overall misclassification cost is then defined as
\begin{equation}
\begin{split}
\mathcal{J}(w) &= \frac{1}{n}\sum_{i=1}^{n}C(h(\mathbf{w}^T\mathbf{d}_i),\ell_i) \\
&= -\frac{1}{n}\sum_{i=1}^{n}\big(\ell_i\log h(\mathbf{w}^T\mathbf{d}_i)  \\
&+ (1-\ell_i)\log(1-h(\mathbf{w}^T\mathbf{d}_i))\big)\;,
\end{split}
\end{equation}
where $n=\omega+\lambda-1$. Our final goal is thus to find the weight vector $\mathbf{w}$ which can minimize $\mathcal{J}(w)$.

We use the gradient descent method \cite{BoydCon01} to solve the above minimization problem. In particular, we update every weight $w_j$ as
\begin{equation}
\begin{split}
w_j &= w_j - \alpha\frac{\partial \mathcal{J}(w)}{\partial w_j} \\
&=w_j - \alpha\sum_{i=1}^{n}(h(\mathbf{w}^T\mathbf{d}_i)-\ell_i)d_{i,j},
\end{split}
\end{equation}
where $\alpha$ and $d_{i,j}$ denote the learning step and the $i$th curve's $j$th DTW distance, respectively. The updating process ends when $\mathcal{J}(w)$ stops decreasing, in which case we obtain the optimal weight vector $\mathbf{w}$.

There are four remarks to make. First, the weight vector will be different for different device owners, which totally depend on his $\omega$ sample curves and the $\lambda$ preloaded random curves. Second, the weight vectors resulting from different reference curves chosen from the $\omega$ sample curves are very similar, which have been confirmed by our experiments as well. Third, if the device owner uses multi-curve authentication, the weight vectors for different curve password are highly likely to be different. Finally, if multiple sample curves are chosen as reference curves, the above minimization process needs to run for each of them.

\subsection{Creation of Authentication Templates}\label{sec:Selection}
The final authentication template includes the feature data of every chosen reference curve and its corresponding weight vector from the \texttt{Classifier Training} module, which is referred to as a curve sub-template. If the device owner uses $M\geq 2$ curve password with $\omega$ sample inputs during the enrollment phase, the authentication template also includes a hand-geometry sub-template. The hand-geometry sub-template is denoted by $[\epsilon^-_i,\epsilon^+_i]$, where $\epsilon^-_i$ and $\epsilon^+_i$ denote the minimum and maximum distances between the $i$th finger pair among the $\omega$ samples of the $M$ curves.

\section{Verification}\label{Sec:Verification}
The verification phase starts when someone tries to unlock the mobile device. The user will be prompted to input his password. The verification phase uses the same \texttt{Touch Monitoring}, \texttt{Data Processing}, and \texttt{Feature \\
Extraction} modules to extract the data for the ten features. The extracted data are then inputted into a \textsf{Verification} module, where three tests are performed.
\begin{itemize}
  \item \textbf{Curve test}: If single-curve authentication is used, the candidate curve is tested using the trained classifier. If it is classified as Class I (or Class II), the curve test succeeds (or fails). If multi-curve authentication is used, every candidate curve in the input need to be tested within the corresponding classifier. If all the candidate curves are classified as Class I, the curve test succeeds and fails otherwise.

  \item \textbf{Hand-geometry test}: This test applies to multi-curve cases only. If the distance between every finger pair in the input falls between the corresponding minimum and maximum distances in the handle-geometry template, this test succeeds and fails otherwise.
\end{itemize}

The user is considered legitimate and allowed in only when his input passes all the tests above; otherwise, he is denied access. Although we have tried to minimize false negatives, an authorized user may be denied access in very rare situations, e.g., due to sudden memory loss. One remedy is to let the authentication template include multiple reference curves and their corresponding classifiers. The curve test is said to succeed if the candidate curve(s) is classified as Class~I by any classifier. As another remedy, TouchIn can be combined with a traditional password-based authentication system which is normally too inconvenient to use and only invoked as the last resort. For example, a visually impaired person may ask his friend to help unlock his mobile device using a super password too difficult for himself to input.
\begin{figure}[t]
\centering
    \subfigure[Classification accuracy] {
        \label{accuracy}
        \includegraphics[width=0.225\textwidth]{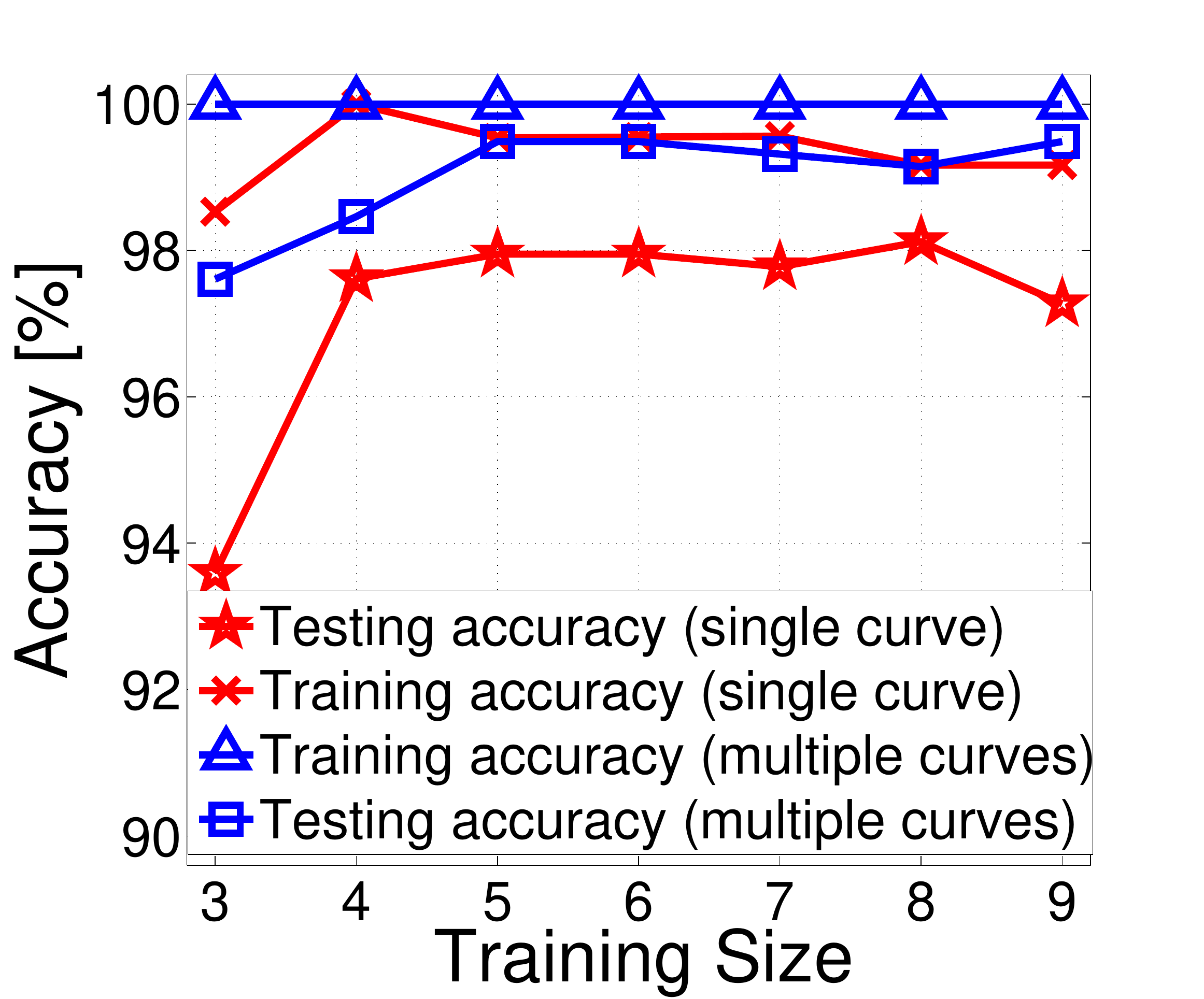}
    }
\hfill
    \subfigure[False-positive/negative rates] {
        \label{rate}
        \includegraphics[width=0.225\textwidth]{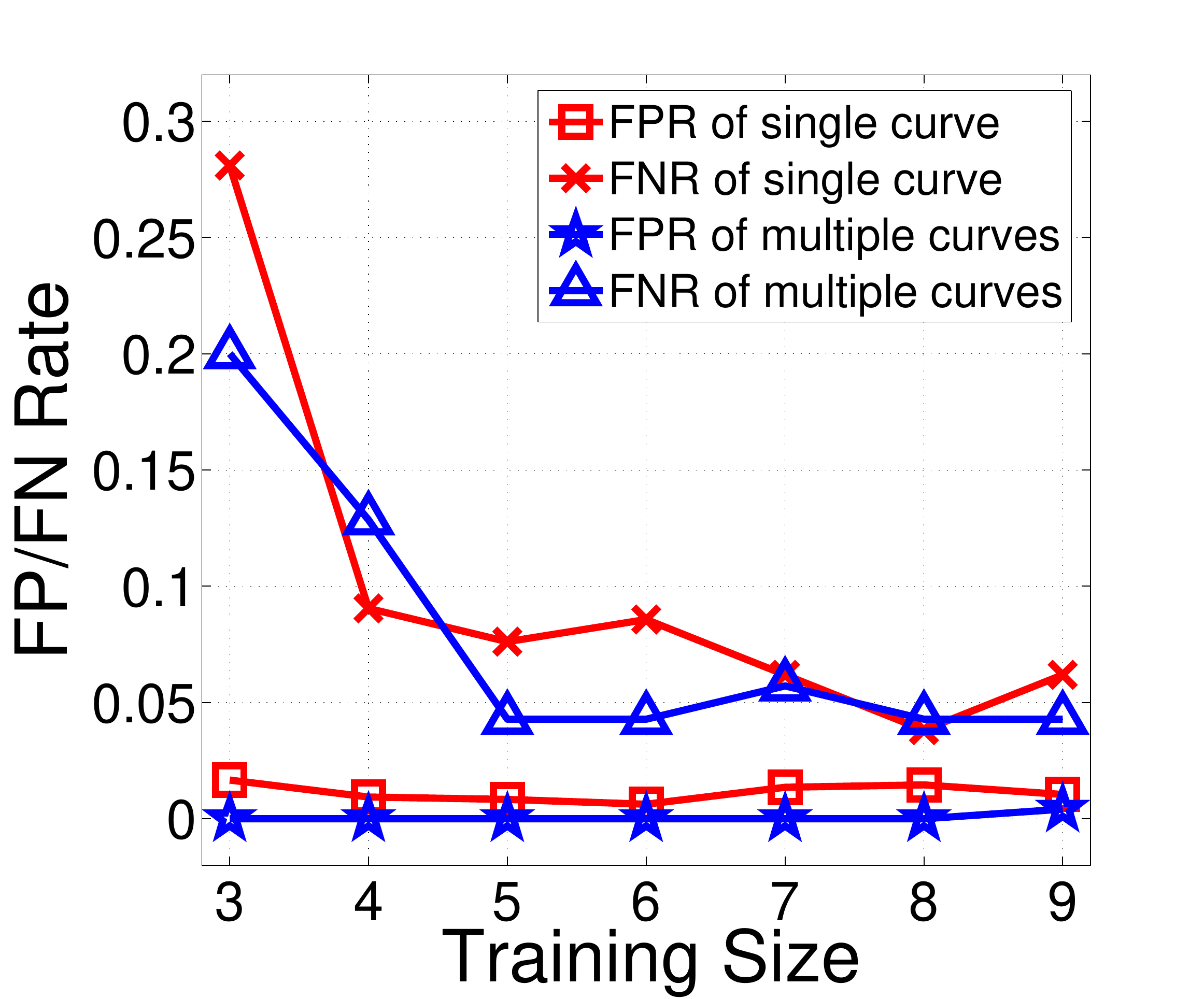}
    }
\caption{Impact of the training-set size.}\label{Trainingsize}
\end{figure}

\section{Performance Evaluation}\label{sec:Experiment}
In this section, we report the experimental security and usability studies of TouchIn on Google Nexus 7 tablets. In what follows, we first introduce our data-collection process and key performance metrics. Then we report the evaluation results for both security and usability features.

\subsection{Data Acquisition}\label{SubSec:DataAcquisition}
We used the same data collection process in Section~\ref{subsec:DataCollection}, in which 3600 single-curve password samples and 3600 multi-curve password samples were collected from 30 volunteers who closed their eyes while drawing on the touchscreen. In addition, we recruited 10 more volunteers with similar background to mimic attackers. We made videos of legitimate volunteers drawing curves on the touchscreen and showed the videos to attackers. Since it is infeasible to ask each attacker to watch too many videos, we let each attacker watch the videos of three randomly selected victims with each drawing one curve password randomly chosen from his/her three. In accordance with the adversary models in Section~\ref{sec:Adversary}, we considered the following attacks.

\begin{itemize}
 \item \textbf{Attack 1: one-time observation}. The attackers were allowed to observe how the victims input their curve password once without knowing other information. More specifically, we first showed the video of each victim's authentication process with one finger to each attacker once and asked each attacker to make five authentication attempts for each victim. Next, we showed the video of each victim's authentication process with two fingers to each attacker once and also asked each attacker to mimic each victim for five times. We collected $3\times5\times10\times2=300$ mimicked curve samples for three victims in total for this attack.

  \item \textbf{Attack 2: four-time observations}. The attackers continued observing the videos of single-finger and two-finger authentication processes of each victim three more times and then produced five mimicked curve samples for each victim. Besides, ten attackers also observed the video of two-finger authentication process of each victim three more times, and produced another five mimicked curve samples. We also collected $3\times5\times10\times2=300$ mimicked curve samples for this attack.

  \item \textbf{Attack 3: four-time observations and rough curve information}. The attackers were additionally provided with the rough shapes of each victim's curve password of the three victims, e.g., whether the curve password is an ellipse or a rectangle. Then each attacker was asked to make five attempts with both one finger and two fingers for each victim, leading to $3\times5\times10\times2=300$ mimicked curve samples for this attack.

  \item \textbf{Attack 4: four-time observations and exact curve information}. The attackers were finally presented with the exact shape of the curve password. Each of them then made five more attempts with both one finger and two fingers for each victim, producing another $3\times5\times10\times2=300$ mimicked curve samples.
\end{itemize}

Attacks 1 and 2 above correspond to the Type-II adversary introduced in Section~\ref{sec:Adversary}, and Attacks 3 and 4 correspond to the Type-III and Type-IV adversaries, respectively. Since Type-I attackers refer to those totally blind to the curve password, we can simply evaluate their impact by testing the collected curve samples of legitimate users against those of others, as to be shown later.
\begin{figure}[t]
\centering
    \subfigure[Precision-Recall curves] {
        \label{3subfigure 1}
        \includegraphics[width=0.225\textwidth]{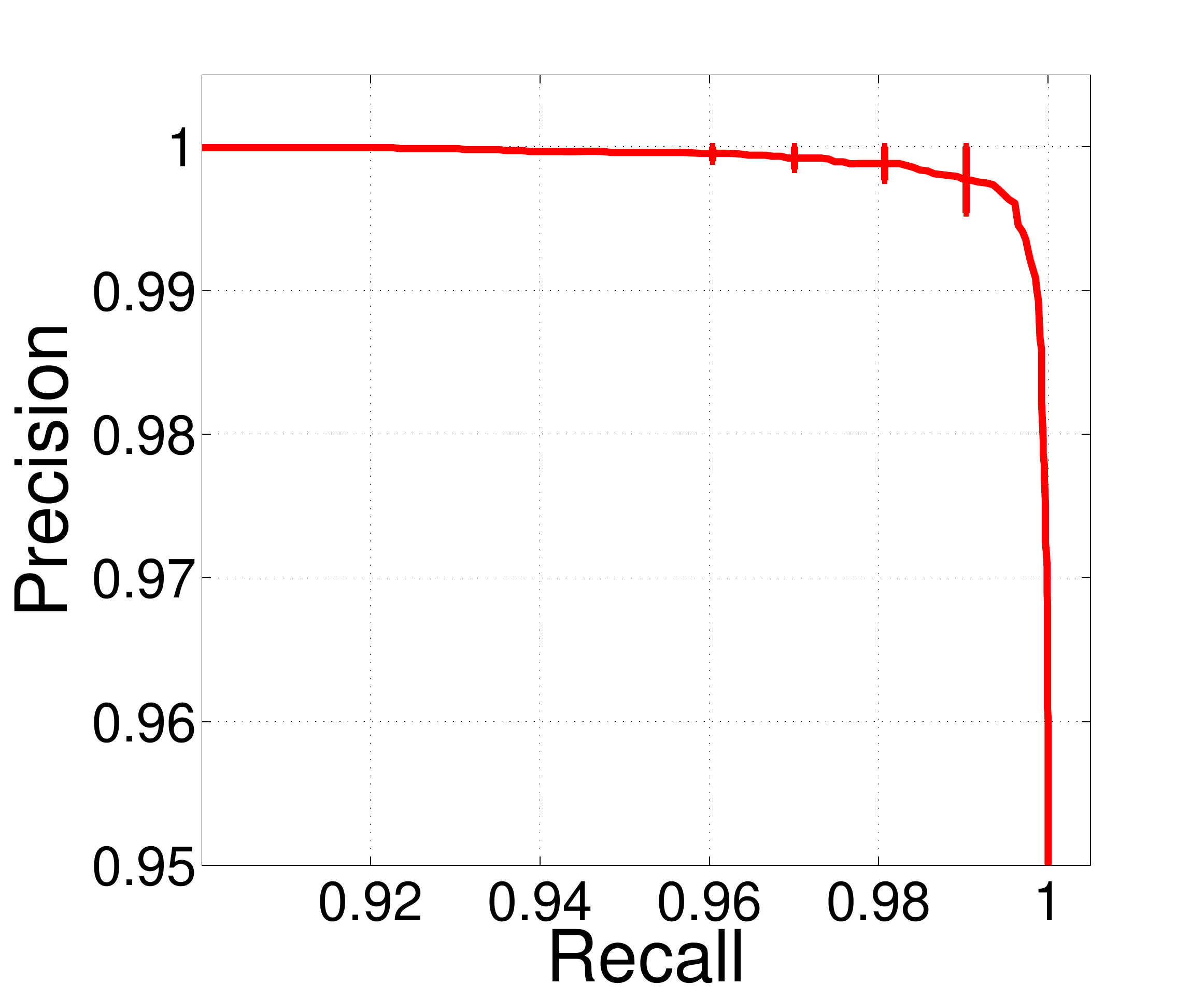}
    }
\hfill
    \subfigure[ROC curves] {
        \label{3subfigure 2}
        \includegraphics[width=0.225\textwidth]{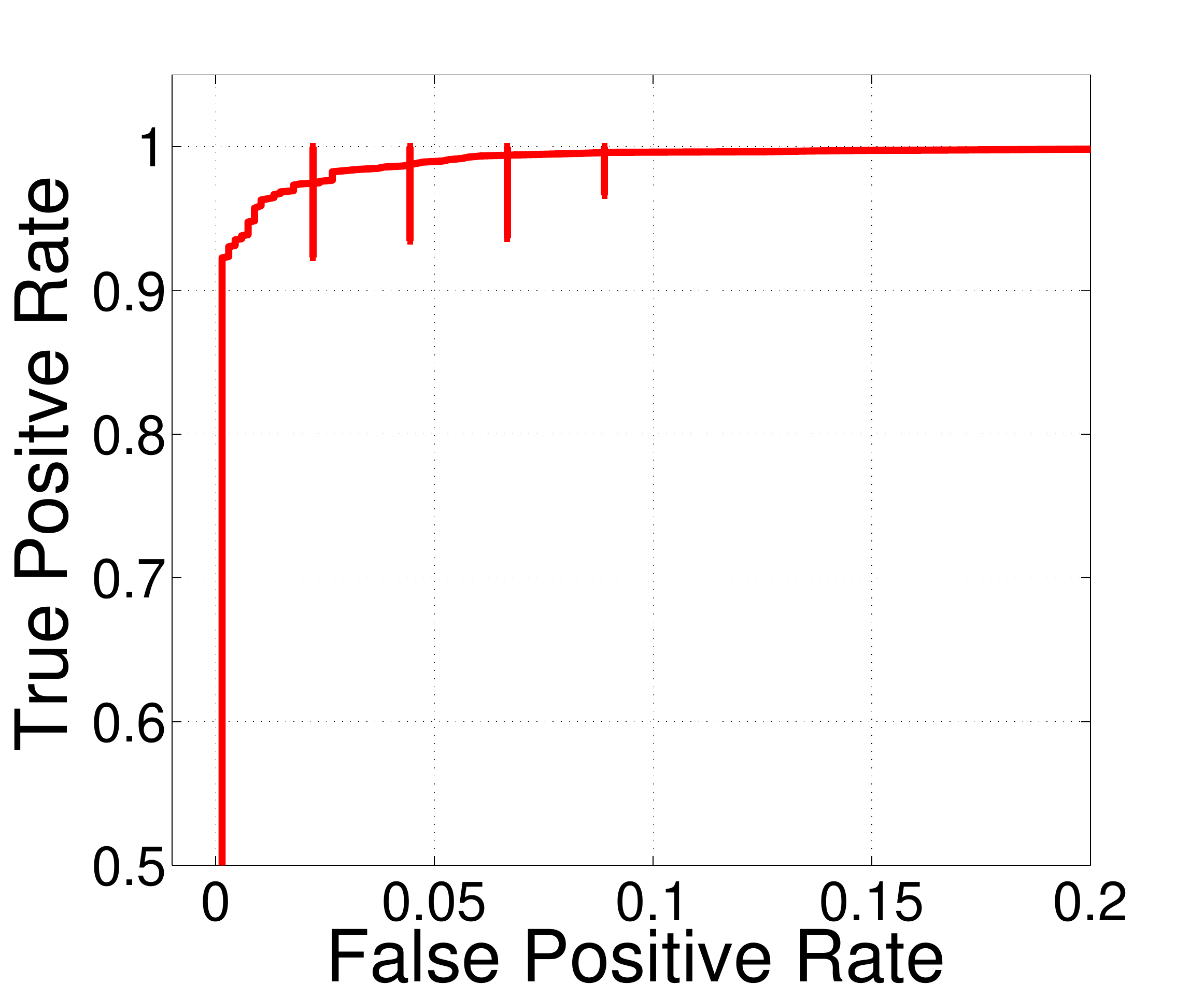}
    }
\caption{Performance of single-curve authentication.}\label{fig3}
\end{figure}

\begin{figure}[t]
\centering
    \subfigure[Precision-Recall curves] {
        \label{4subfigure 1}
        \includegraphics[width=0.225\textwidth]{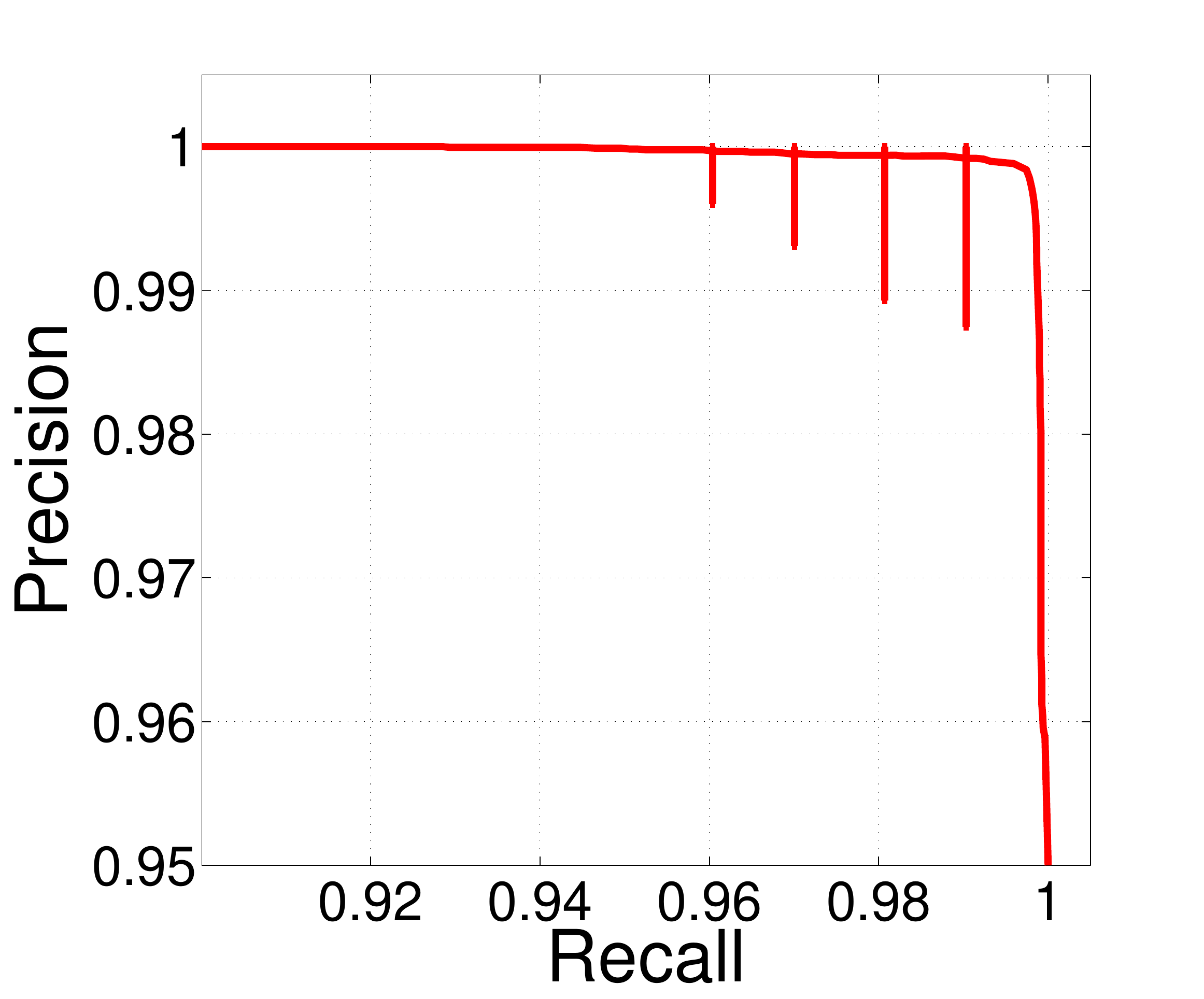}
    }
\hfill
    \subfigure[ROC curves] {
        \label{4subfigure 2}
        \includegraphics[width=0.225\textwidth]{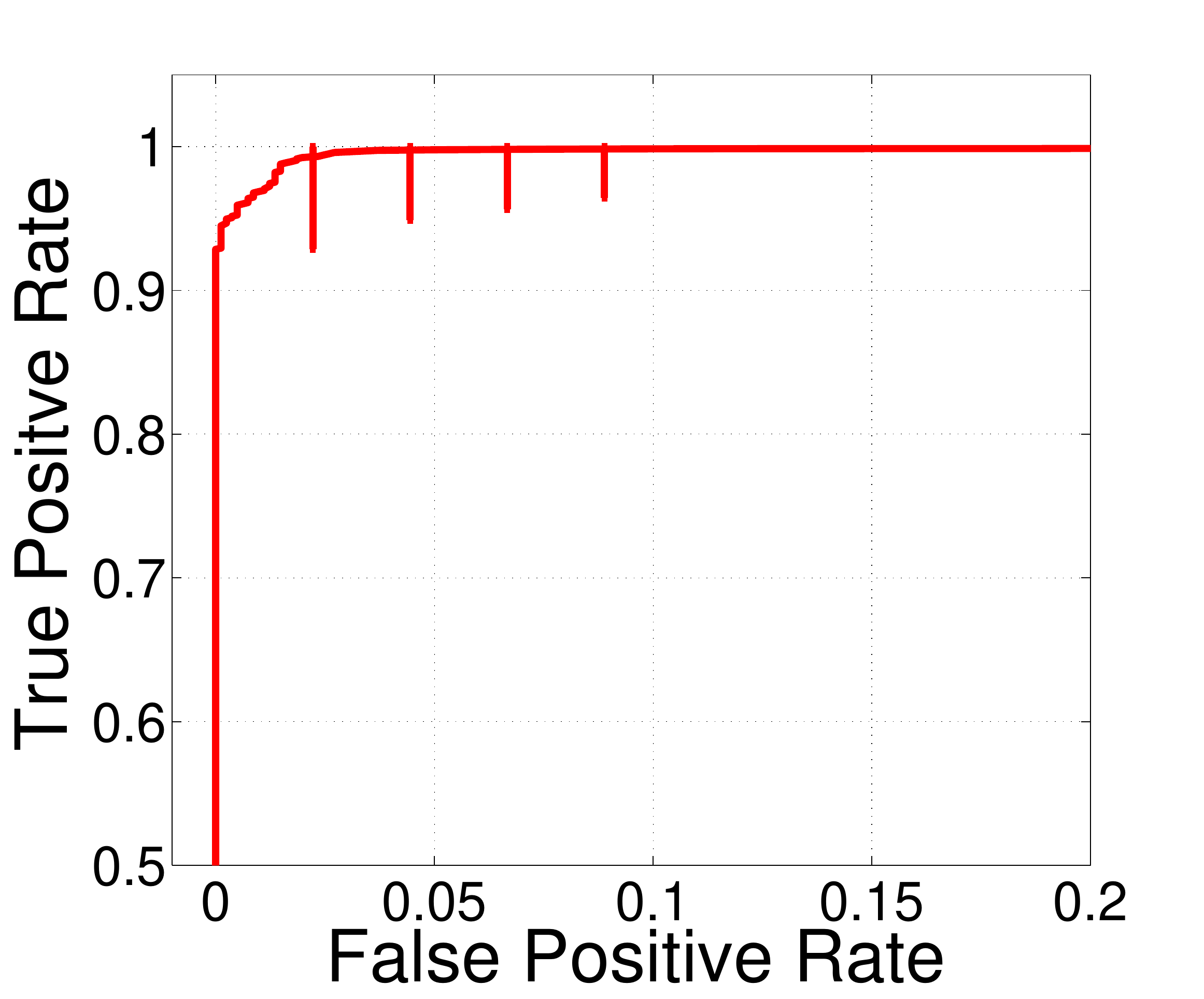}
    }
\caption{Performance of multi-curve authentication.}\label{fig4}
\end{figure}

\subsection{Performance Metrics}
We adopt the following performance metrics.
\begin{itemize}
  \item \textbf{ROC Curve}. In signal detection theory, a receiver operating characteristic curve, or simply ROC curve, is a graphical plot commonly used to illustrate the performance of a binary classifier as its discrimination threshold is varied. A ROC curve is created by plotting the true-positive rate versus the false-positive rate. More specifically, let $\#_{\mathsf{TP}}$, $\#_{\mathsf{FP}}$, $\#_{\mathsf{TN}}$, and $\#_{\mathsf{FN}}$ denote the number of true positive, false positive, true negative and false negative, respectively. The true-positive and false-positive rates are calculated respectively as
      \begin{equation}
      \mathtt{TPR} = \frac{\#_{\mathsf{TP}}}{\#_{\mathsf{TP}} + \#_{\mathsf{FN}}}
      \end{equation}
      and
      \begin{equation}
      \mathtt{FPR} =  \frac{\#_{\mathsf{FP}}}{\#_{\mathsf{FP}} + \#_{\mathsf{TN}}}.
      \end{equation}
      TouchIn should achieve both a high true-positive rate and a low false-positive rate.

  \item \textbf{Precision-Recall Curve}. In our scenario, precision is the measure of accuracy provided by the authentication system. It represents the percentage of legitimate users out of those passing the authentication and can be computed as
       \begin{equation}
      \mathtt{Precision} = \frac{\#_{\mathsf{TP}}}{\#_{\mathsf{TP}} + \#_{\mathsf{FP}}}.
      \end{equation}
      Recall is the measure of the capability that the authentication system can pass authorized users and is defined as $\mathtt{Recall}=\mathtt{TPR}$.

  \item \textbf{Authentication Time}. This refers to the time for the mobile device to decide whether to allow the user in and should be as short as possible.
\end{itemize}

\subsection{Experimental Results}
In this section, we present our experimental results to show the performance of our authentication system.
\begin{figure}[t]
\centering
    \subfigure[Precision-Recall curves] {
        \label{attsubfigure 1}
        \includegraphics[width=0.34\textwidth]{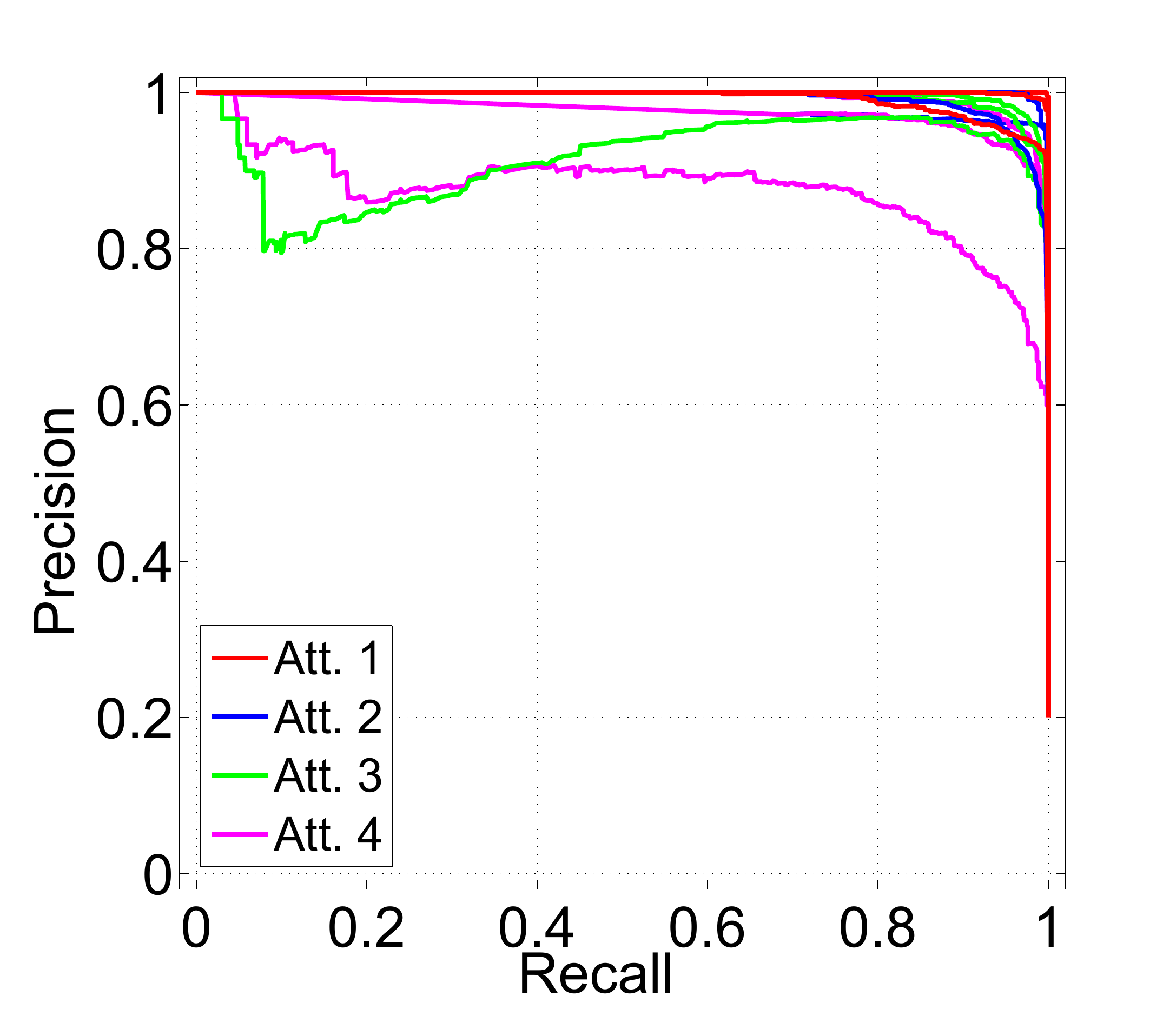}
    }
\hfill
    \subfigure[ROC curves] {
        \label{attsubfigure 2}
        \includegraphics[width=0.34\textwidth]{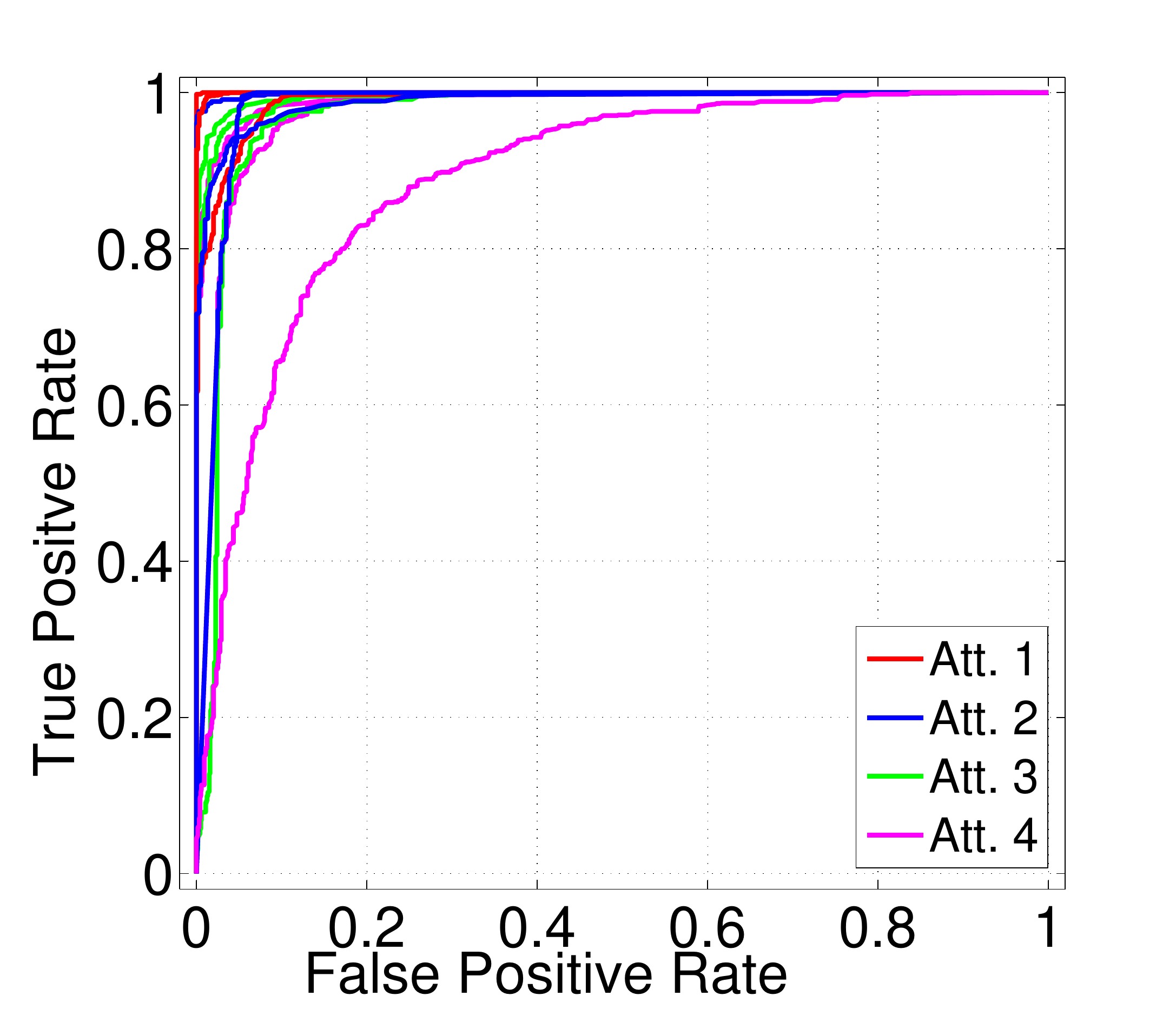}
    }
\caption{Single-curve authentication under attacks.}\label{att1}
\end{figure}

\begin{figure}[t]
\centering
    \subfigure[100\% precision] {
        \label{bar1subfigure 1}
        \includegraphics[width=0.225\textwidth]{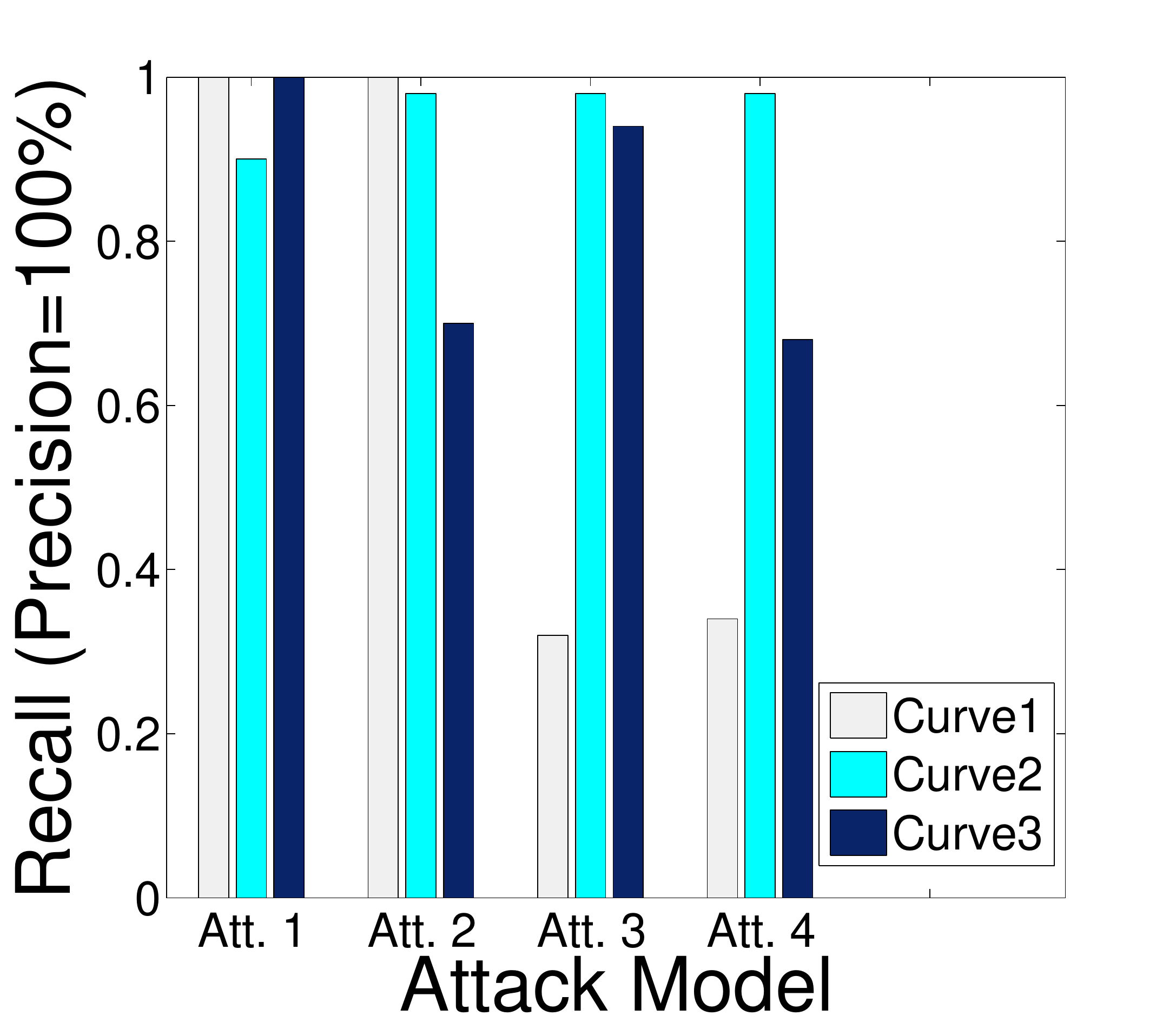}
    }
\hfill
    \subfigure[75\% recall] {
        \label{bar1subfigure 2}
        \includegraphics[width=0.225\textwidth]{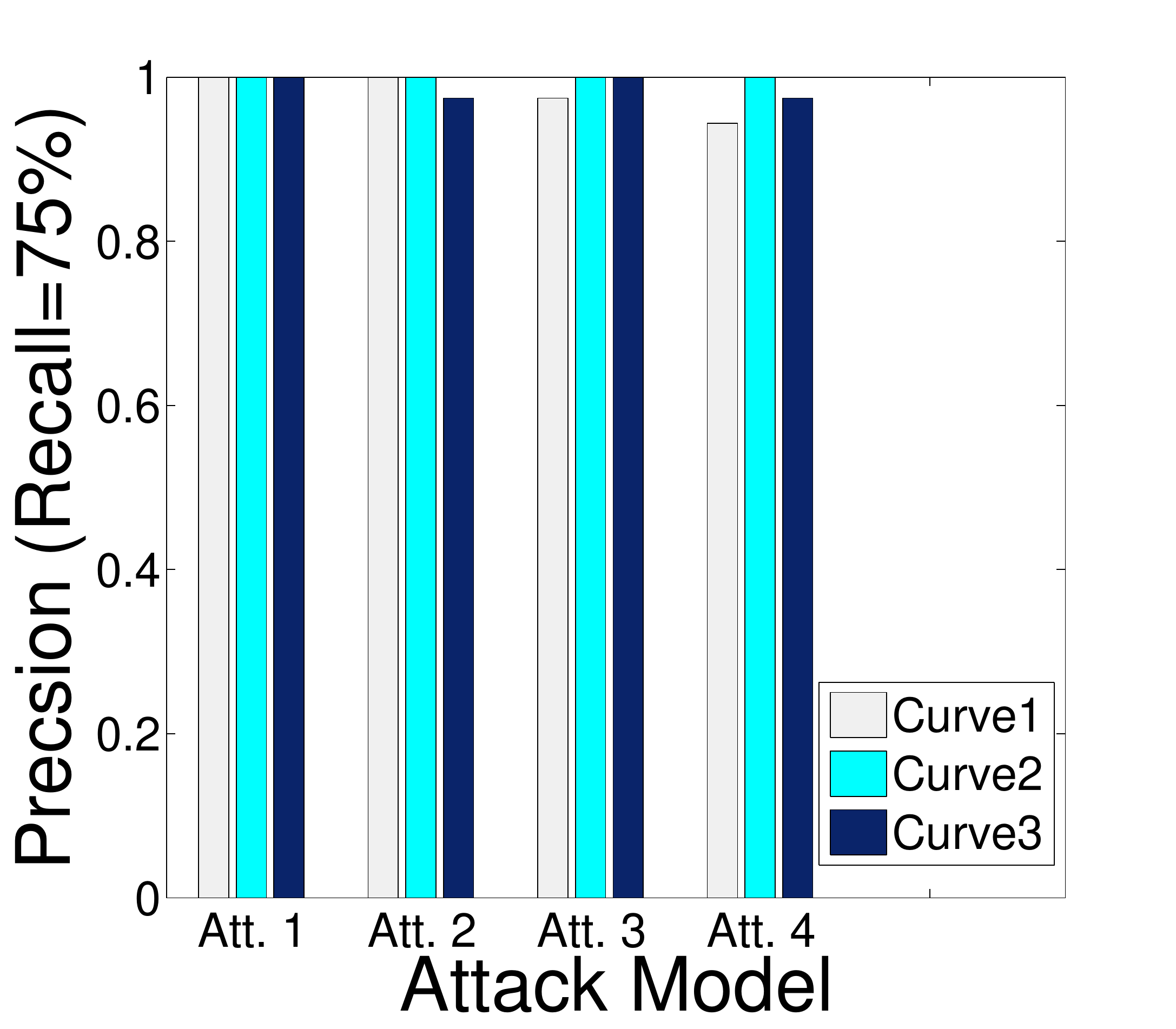}
    }
\caption{Achievable recall and precision under single-curve authentication.}\label{bar1}
\end{figure}
\vspace{0.2in}

\subsubsection{Performance for Legitimate Authentication}
We first demonstrate our system performance for legitimate users only. Recall that each of the 30 volunteers chose three curve password for single-curve authentication and input each curve 20 to 50 times, producing 3,600 single-curve samples. Then we randomly chose $\omega$ samples for each of the $30\times 3$ curve password to form a training set of $90\omega$ sample curves for the classifier. This corresponds to having $\lambda=89\omega$ preloaded random curves for each curve password. The rest curve samples composed a testing set. We did the same for the 3,600 multi-curve samples. We conducted the evaluation for every legitimate user each with three different curve passwords, and each reported data point represents the average of 90 authentication instances, unless otherwise stated.

\vspace{.1in}\noindent\textbf{\underline{Impact of the Training-set Size:\;}} We first show the impact of the training-set size on the classification result. Intuitively, the larger the training set, the more information provided, the better the classifier performance. We varied the size of the training set by changing $\omega$ and recorded the corresponding training-accuracy measurements, testing-accuracy measurements, the false-positive rate, and the false-negative rate.

Fig.~\ref{accuracy} shows the impact of the training-set size on training accuracy and testing accuracy of single-curve authentication and multi-curve authentication, respectively, where the training (or testing) accuracy is defined as the ratio of true positives and negatives to the total number of curve samples involved in the training (or testing) process. We can see that the classification accuracy increases as $\omega$ increases and stays very high as $\omega\geq 4$. In addition, we can see that the false-positive and false-negative rates of both single-curve authentication and multi-curve authentication decrease as $\omega$ increases and stay sufficient low as $\omega\geq 5$. Such results are desirable because $\omega$ is directionally proportional to the enrollment time and thus usability.

\vspace{.1in}\noindent\textbf{\underline{Authentication Performance:\;}} Now we show the authentication performance of TouchIn. For each curve password of every legitimate user, its remaining samples in the testing set can be regarded as the user's inputs at different times, while all the other samples in the testing set can be regarded as the inputs by a Type-I attacker. For clarity and simplicity, we report the average results as well as the upper and lower bounds for all the curve passwords in the precision-recall and ROC curves.

Fig.~\ref{fig3} illustrates the performance of single-curve authentication. We can see that the average precision-recall curve is close to the top-right corner, which indicates that our system can obtain high precision and high recall simultaneously. Similarly, the average ROC curve is close to the top-left corner, which indicates that our system can achieve a high true-positive rate together with a low false-positive rate. Both precision-recall and ROC curves show that our system is good at distinguishing legitimate and illegitimate users.

Fig.~\ref{fig4} demonstrates the performance of multi-curve authentication. Similar to single-curve authentication, our system achieves high precision and high recall simultaneously as well as a high true-positive rate and a low false-positive rate. In addition, we can see that the average performance of multi-curve authentication is better than single-curve authentication. This is anticipated because more information is incorporated in the authentication process.

\subsubsection{Performance Under Attacks}
In this part, we show the system performance under the four attacks defined in Section~\ref{SubSec:DataAcquisition}.

We first report the performance of single-curve authentication under attacks. For every attack and each curve password, we ran the experiments 30 times to obtain the average precision-recall curves in Fig.~\ref{attsubfigure 1} and ROC curves in Fig.~\ref{attsubfigure 2}, in which every attack is associated with three identical-colored curves with each corresponding to one victim. From Fig.~\ref{att1}, we can see that as the attack strength increases, the performance of our system decreases. This is anticipated because the attackers know more information about the curve password as the attack strength increases.

\begin{figure}[t]
\centering
    \subfigure[Precision-Recall curves] {
        \label{att2subfigure 1}
        \includegraphics[width=0.34\textwidth]{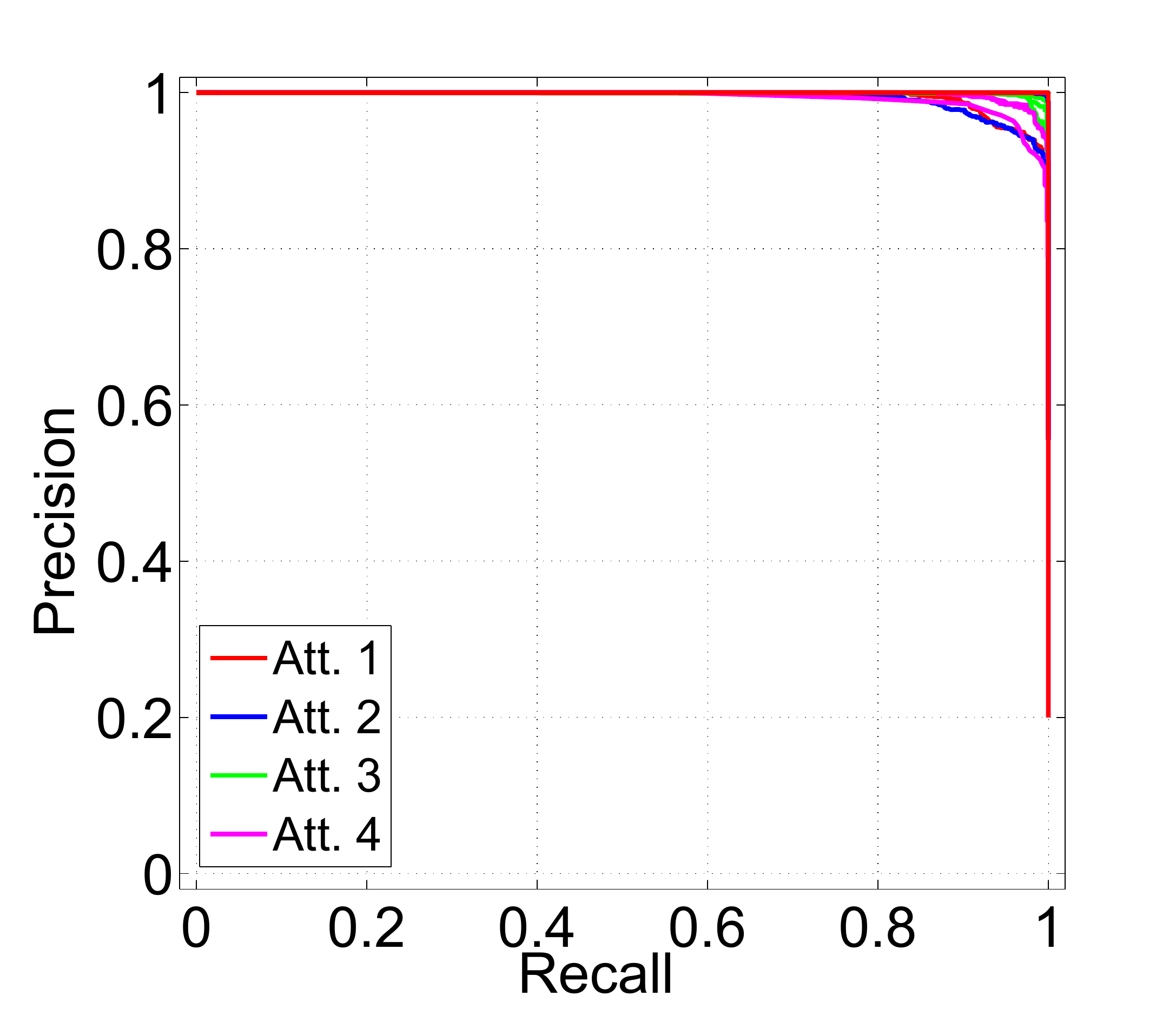}
    }
\hfill
    \subfigure[ROC curves] {
        \label{att2subfigure 2}
        \includegraphics[width=0.34\textwidth]{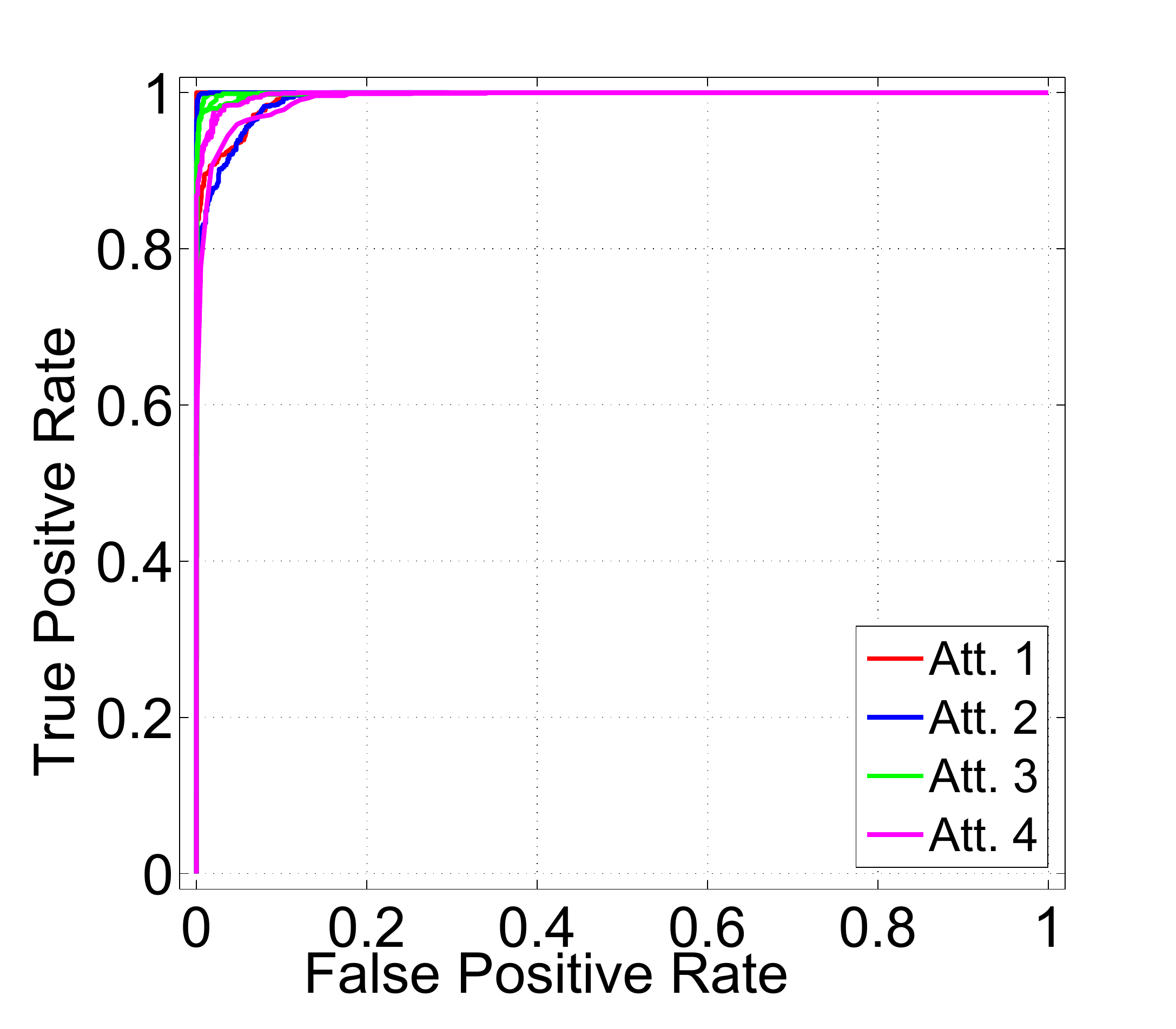}
    }
\caption{Multi-curve authentication under attacks.}\label{att2}
\end{figure}

\begin{figure}[t]
\centering
    \subfigure[100\% precision] {
        \label{bar2subfigure 1}
        \includegraphics[width=0.225\textwidth]{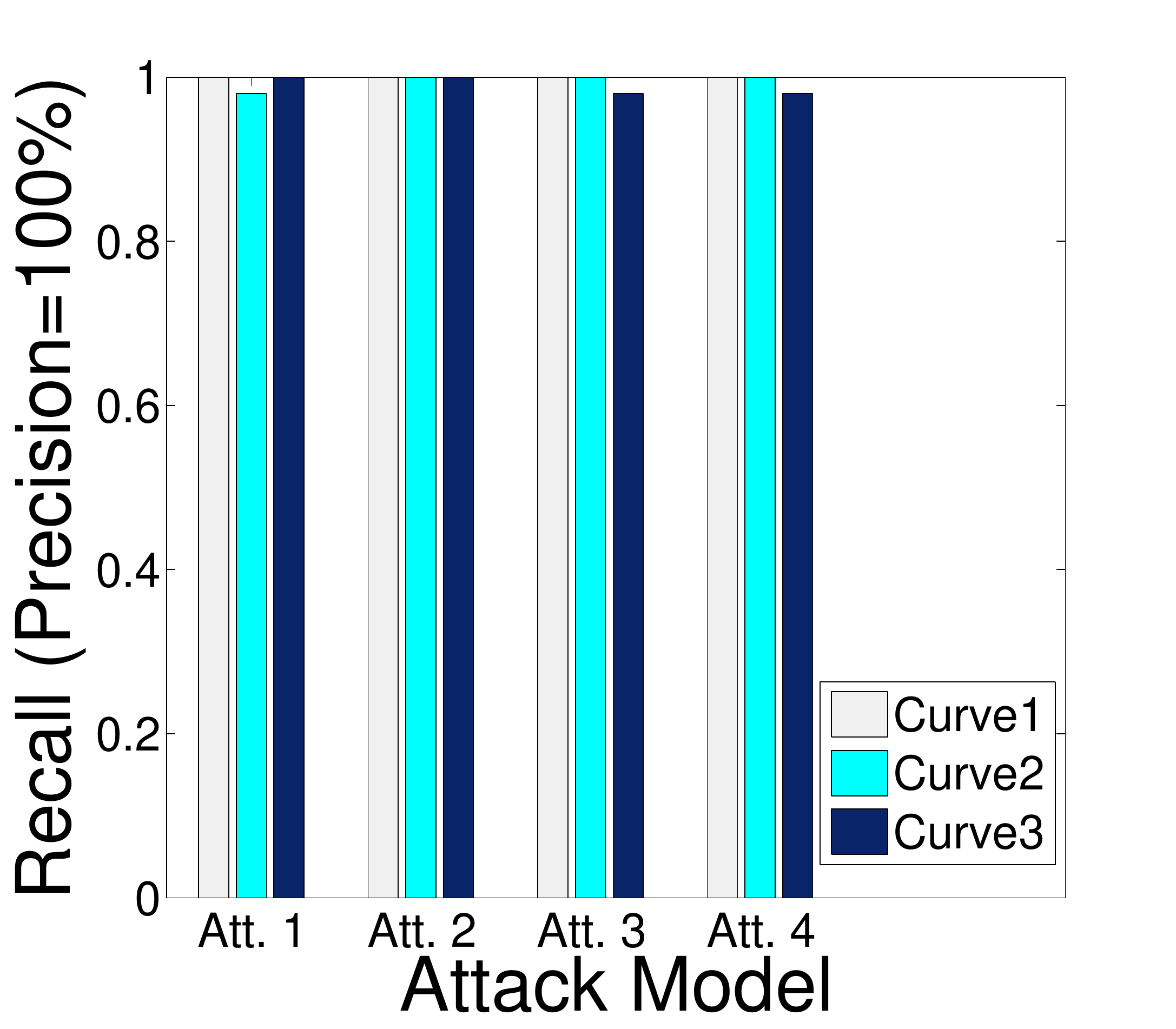}
    }
\hfill
    \subfigure[75\% recall] {
        \label{bar2subfigure 2}
        \includegraphics[width=0.225\textwidth]{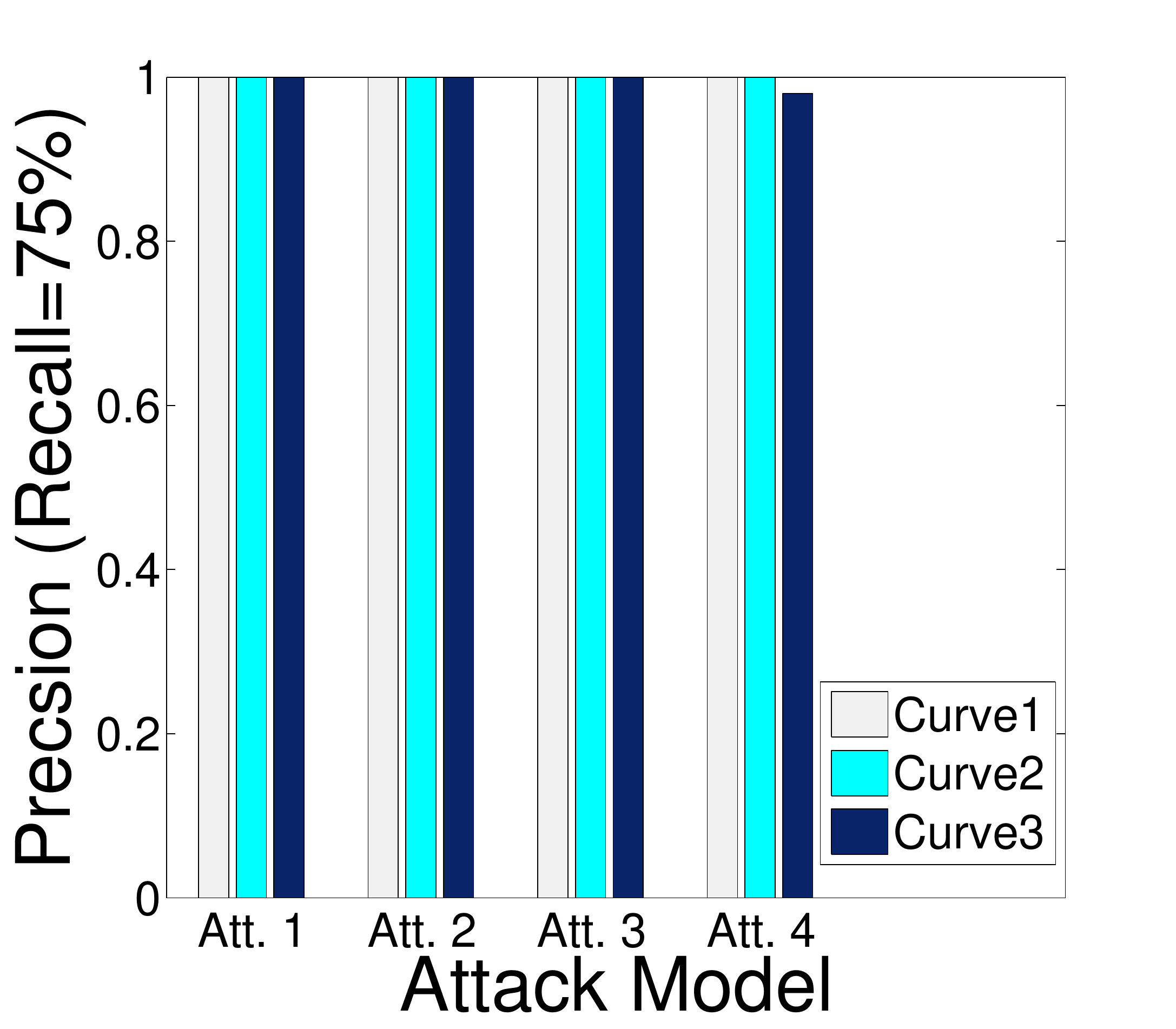}
    }
\caption{Achievable recall and precision under multi-curve authentication.}\label{bar2}
\end{figure}

In Fig.~\ref{bar1subfigure 1}, we show the maximum achievable recall for each curve password under the four attacks, where the precision is required to be 100\%. It illustrates the trade-off between security and usability. If a user wants better usability, i.e., fewer tries needed on average to pass the authentication, then the system configuration will result in a higher recall rate. On the other hand, it will be difficult to deal with more challenging attackers. The maximum precision with a 75\% recall rate is shown in Fig.~\ref{bar1subfigure 2}.  It illustrates how likely an attacker can pass the authentication by a fixed number of attempts on average. We can see that our system can reject all the attacks with an over 90\% probability on average.

Next, we show the performance of multi-curve authentication under the attacks. The similar trend can be found in Fig.~\ref{att2} and Fig.~\ref{bar2}. In particular, the system performance decreases as the attack strength increases. In addition, it is obvious that multi-curve authentication performs better than single-curve authentication under various attack models. The reason is that multi-curve authentication compares the features of multiple corresponding curves and also checks other feature such as hand geometry, which are much more difficult to infer by the attackers.

\subsubsection{Computation Overhead}
The computation overhead of TouchIn lies in two aspects. The first is mainly incurred by classifier training in the enrollment phase, and the second is mainly caused by feature extraction, DTW distance calculations, and classification in the verification phase. The computation overhead should be sufficiently low for fast enrollment and verification.

\begin{figure}
\centering
    \subfigure[Enrollment time.] {
        \label{fig:train}
        \includegraphics[width=0.225\textwidth]{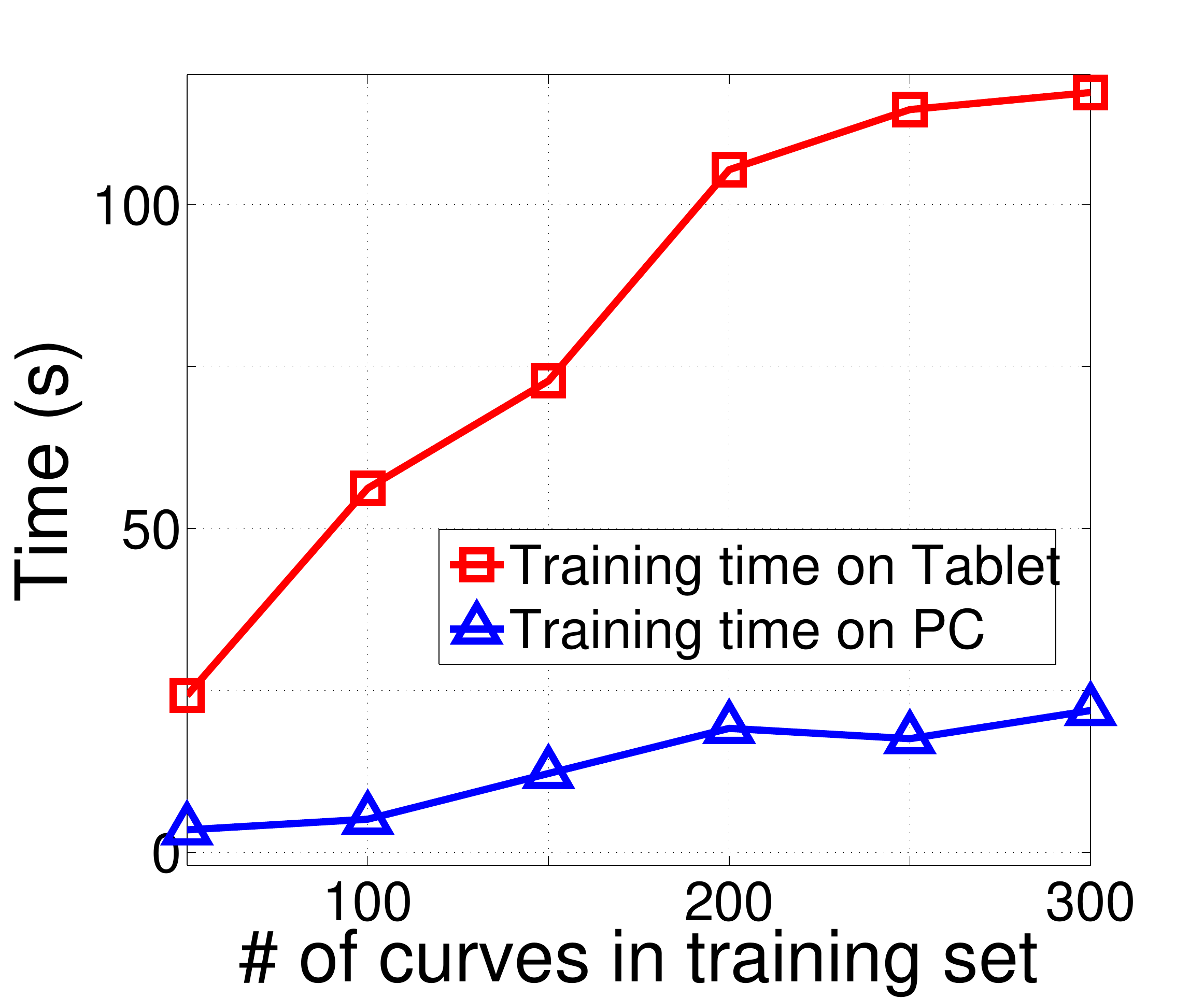}
    }
\hfill
    \subfigure[Verification time.] {
        \label{fig:dtw}
        \includegraphics[width=0.225\textwidth]{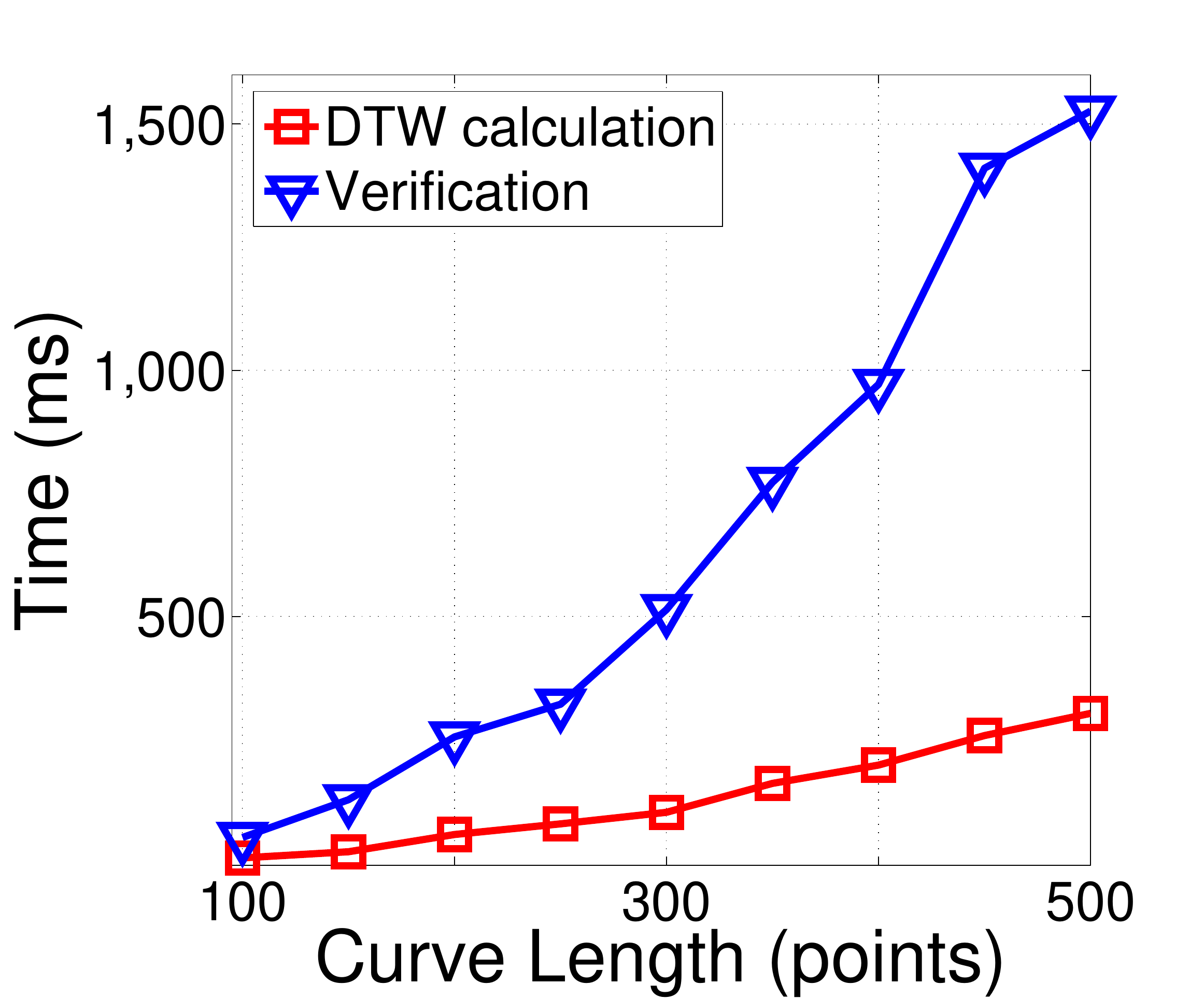}
    }
\caption{Computation overhead for single-curve authentication.}\label{computationoverhead}\vspace{-0.2in}
\end{figure}

Classifier training can be done either on the mobile device or through a trusted third party as in \cite{LiUno13}. Fig.~\ref{fig:train} shows the enrollment time for single-curve authentication when classifier training is done on a Google Nexus 7 tablet and a Dell desktop with 2.67 GHz CPU, 9 GB RAM, and Windows 7 64-bit Professional. We can see that the enrollment time on Google Nexus 7 is about 120 seconds when the training set contains 300 curve samples. Although such enrollment time seems long, it is still acceptable because classifier training is a one-time process and can be done when the user does not use the mobile device. As in \cite{LiUno13}, an alternative way for classifier training is to outsource it to a trusted third-party server computing and returning the weight vector to the user. As shown in Fig.~\ref{fig:train}, the enrollment time on the Dell desktop is much shorter than that on Google Nexus 7 and  can be much more shortened on a more powerful cloud server. The enrollment time for multi-curve authentication involving $M$ curves is approximately $M$ times that for multi-curve authentication.

The verification time for single-curve authentication is shown in Fig.~\ref{fig:dtw}. According to our experiments on Google Nexus 7 tablet with Android 4.2, the time for feature extraction and classification is less than one millisecond and thus negligible. In contrast, the time for DTW distance calculations increases with the number of feature points on an curve password. To shed some light on the verification overhead, Fig.~\ref{fig:dtw} shows the average time for single-feature DTW distance calculations and the overall verification time which involves the DTW distance calculations for nine curve features. As we can see, the average DTW calculation time increases as the number of curve points increases. According to our experimental data, most curve passwords contain less than 300 points which only cost less than 100 ms for single-feature DTW distance calculation, and the overall verification time is below 900 ms. As to multi-curve authentication with $M$ curves, the verification time is approximately $M$ times the single-curve verification time.

\vspace{-.15in}
\begin{table}[ch]
\caption{Comparison with existing schemes}
\begin{tabular}{|c|c|c|c|c|}
\hline
\multirow{2}{*}{Scheme} & \multicolumn{2}{|c|}{TPR} & \multicolumn{2}{|c|}{FPR} \\
\cline{2-5}
& Single & Multiple & Single & Multiple\\
\hline
TouchIn & 97.5\% & 99.3\% & 2.3\% & 2.2\%\\
\hline
PassChords \cite{AzenkPas12} & \multicolumn{2}{|c|}{83.7\%} & \multicolumn{2}{|c|}{NA} \\
\hline
GEAT \cite{ShahzSec13} & \multicolumn{2}{|c|}{94.6\%} & \multicolumn{2}{|c|}{4.02\%} \\
\hline
\end{tabular}\label{compare}\vspace{-.15in}
\end{table}

\subsubsection{Comparison with Existing Schemes}
Now we compare TouchIn with PassChords \cite{AzenkPas12}, the only known work dedicated to sightless mobile authentication. As shown in Table~\ref{compare}, TouchIn has a much higher TPR (i.e., authentication success rate) than PassChords. In addition, TouchIn has a sufficiently low FPR, and the FPR information is not available in \cite{AzenkPas12}. We also observe that PassChords is vulnerable to shoulder-surfing attacks. To confirm our observation, we asked 10 attackers to observe the PassChords authentication process once, twice, three times, and four times, respectively. 8 out of 10 attackers could successively repeat the PassChords password after just one observation, and the rest two could succeed after two observations. In contrast, TouchIn is highly resilient to shoulder-surfing attacks, as we have shown in Fig.~\ref{att1} and Fig.~\ref{att2}.

We also compare TouchIn with GEAT \cite{ShahzSec13}, a very recent mobile authentication scheme. We can see that TouchIn has slightly better TPR and FPR performance than GEAT, but GEAT does not have the same sightless feature of TouchIn.

\begin{table}[thp]
\centering
\caption{Usability scores}
\begin{tabular}{|c|c|c|c|c|c|}
\hline
  & Mean & Standard Deviation & Min & Median & Max  \\
\hline
Q1 & 4.5 & 0.60698 & 3 & 5 & 5\\
\hline
Q2 & 4.45 & 0.60481 & 3 & 4.5 & 5 \\
\hline
Q3 & 4.4 & 0.68056 & 3 & 4.5 & 5 \\
\hline
Q4 & 4.6 & 0.59824 & 3 & 5 & 5 \\
\hline
Q5 & 4.35 & 0.67.82 & 3 & 4 & 5 \\
\hline
\end{tabular}\label{usability}\vspace{-.2in}
\end{table}

\subsection{Usability Study}
We also evaluated the usability of TouchIn by surveying the experiment volunteers. In particular, we asked them whether TouchIn is easy to use (Q1), whether curve passwords are easy to memorize (Q2), whether TouchIn is faster than the PIN method and Android pattern lock (Q3), whether TouchIn would be easier to use with more practice (Q4), and their preference of TouchIn (Q5) over the PIN method and Android pattern lock. The statistic informative of survey scores are listed in Table~\ref{usability}, where the scores range from one (lowest) to five (highest). The results indicate that TouchIn is very easy to use and more preferable than the PIN method and Android pattern lock.
\vspace{-0.1in}

\section{Conclusion}\label{sec:Conclusion}
In this paper, we presented and evaluated TouchIn, a novel sightless two-factor authentication system on multi-touch mobile device. The high security, efficiency, and usability were confirmed by detailed experiments on Google Nexus 7 tablets. Our future work includes incorporating more behaviorial characteristics and physiological characteristics as well as the something-you-have paradigm into our authentication system. We also plan to formally analyze the security strength of curve-based passwords in TouchIn.

\vspace{-0.1in}

\bibliographystyle{IEEETran}
\bibliography{./wins,./BA,./winsPub}

\end{document}